\documentclass{article}

\usepackage{microtype}                 
\PassOptionsToPackage{warn}{textcomp}  
\usepackage{textcomp}                  
\usepackage{mathptmx}                  
\usepackage{times}                     
\usepackage{cite}                      
\usepackage{tabu}                      
\usepackage{booktabs}                  
\usepackage{color}
\usepackage{url}

\def\autoref#1{\ref{#1}}

\title{Real-Time Visualization in Non-Isotropic Geometries}

\author{Eryk Kopczy\'nski and Dorota Celi\'nska-Kopczy\'nska}

\usepackage{graphicx}
\usepackage{amsmath}
\usepackage{amsfonts}

\def\bbA{\mathbb{A}}
\def\bbB{\mathbb{B}}
\def\bbR{\mathbb{R}}
\def\bbZ{\mathbb{Z}}
\def\bbH{\mathbb{H}}
\def\bbS{\mathbb{S}}
\def\bbE{\mathbb{E}}

\def\ra{\rightarrow}

\def\PSL{{\bf PSL}}
\def\Sol{{\bf Solv}}
\def\Nil{{\bf Nil}}

\def\longonly#1{#1}
\def\shortonly#1{}

\newtheorem{definition}{Definition}

\begin{document}

\maketitle

\begin{abstract}
Non-isotropic geometries are of interest to low-dimensional topologists, physicists and cosmologists. At the same time, they are challenging to comprehend and visualize. 
  We present novel methods of computing real-time native geodesic rendering of non-isotropic geometries.
  Our methods can be applied not only to visualization. They are also essential for potential applications in machine learning and video games.
\end{abstract}

\section{Introduction}



Non-isotropic geometries do not behave the same in all directions. Although, they are less famous than the isotropic geometries, they arise in Thurston's famous geometrization conjecture \cite{thurston1982}.
This conjecture generalizes the Poincar\'e conjecture, one of the most important conjectures in mathematics,  proven by Perelman \cite{perelman}. Every two-dimensional compact manifold can be given a spherical $\mathbb{S}^2$ ,
Euclidean, or hyperbolic geometry $\bbH^2$; the Thurston conjecture states that every three-dimensional compact manifold can be similarly
decomposed into subsets, each of which admitting one of eight geometries, called the Thurston geometries.
The eight geometries include the three isotropic geometries mentioned, two product geometries ($\bbS^2\times\bbR$, $\bbH^2\times\bbR$,
also called $\bbS^2\times\bbE$ and $\bbH^2\times\bbE$),
and three other geometries: {\bf Solv}, {\bf Nil} (twisted $\bbE^2\times\bbR$), and twisted $\bbH^2\times\bbR$ (also called the universal cover of $SL(2, \bbR)$). The interest in {\bf Solv} and {\bf Nil} ranges from
low-dimensional topologists,
geometric group theorists\longonly{, as those geometries exhibit growth patterns typical to solvable and nilpotent groups }\cite{ggtbook}, to physicists \cite{thurston_physics},
and cosmologists, as possible geometries of our Universe  \cite{weeks2001shape}.
Note that not all three-dimensional geometries are Thurston geometries. There are also non-isotropic geometries for which there are no compact manifolds which admit these geometries.

\begin{figure*}[ht]
\centering
\includegraphics[width=0.22\linewidth]{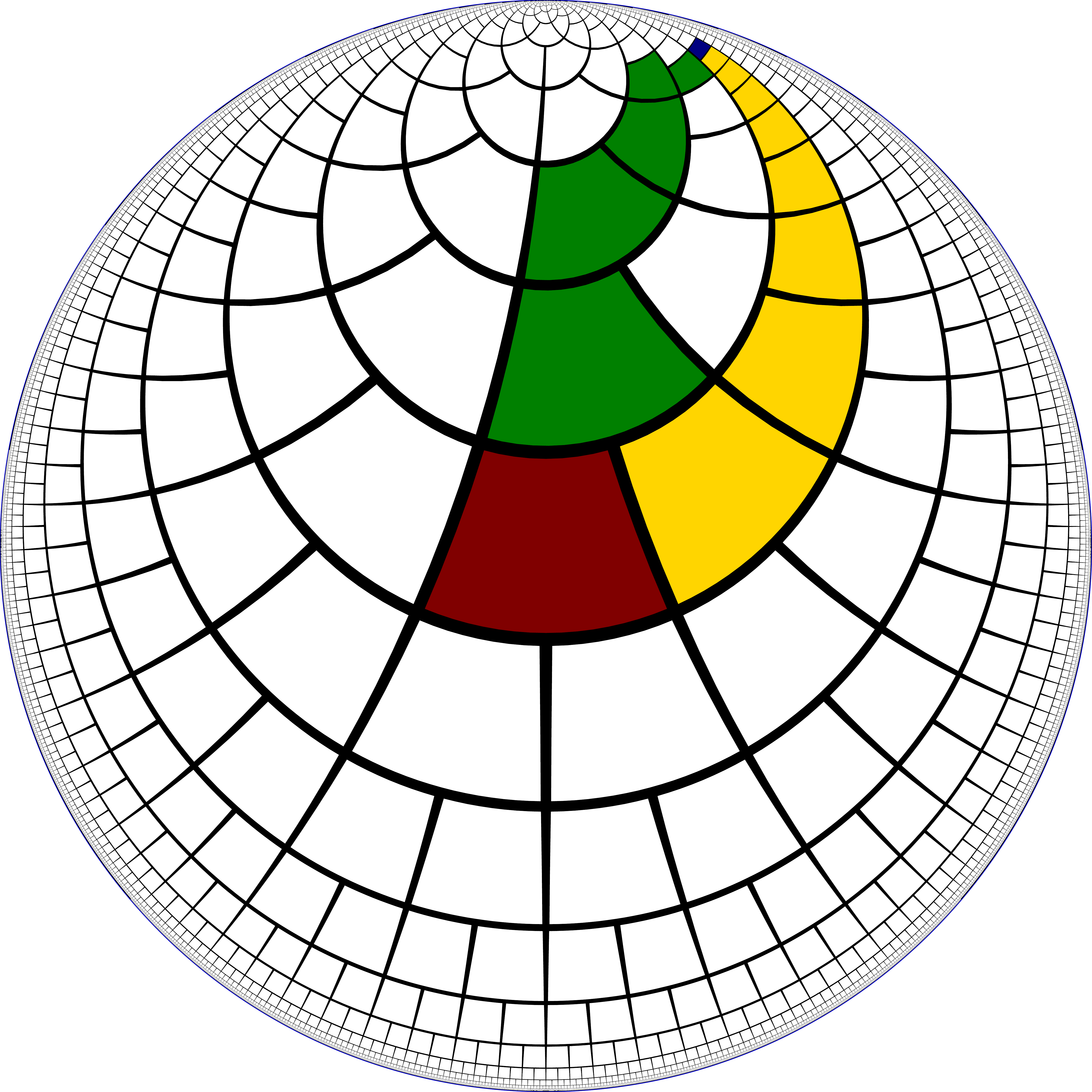}
\includegraphics[width=0.22\linewidth]{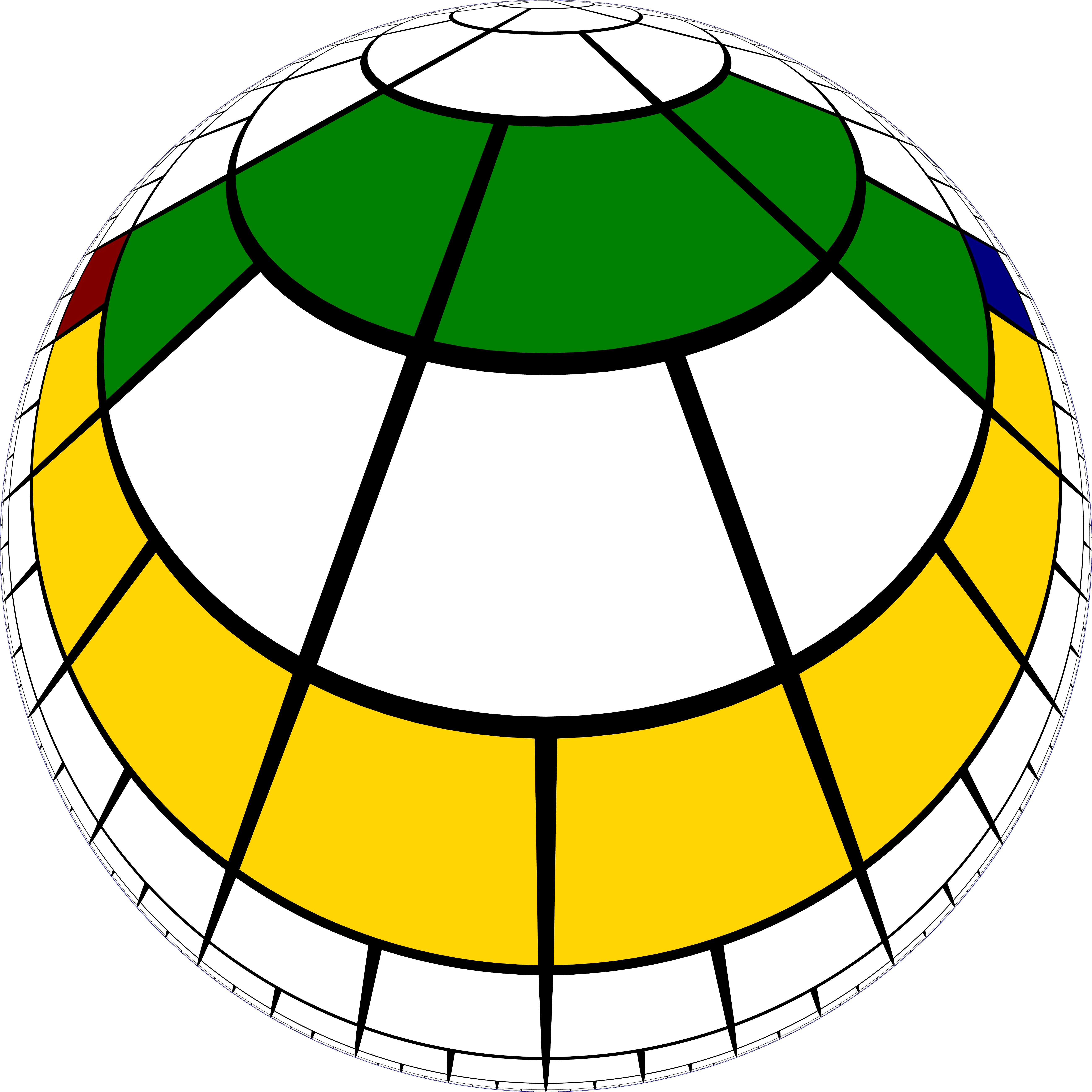}
\includegraphics[width=0.22\linewidth]{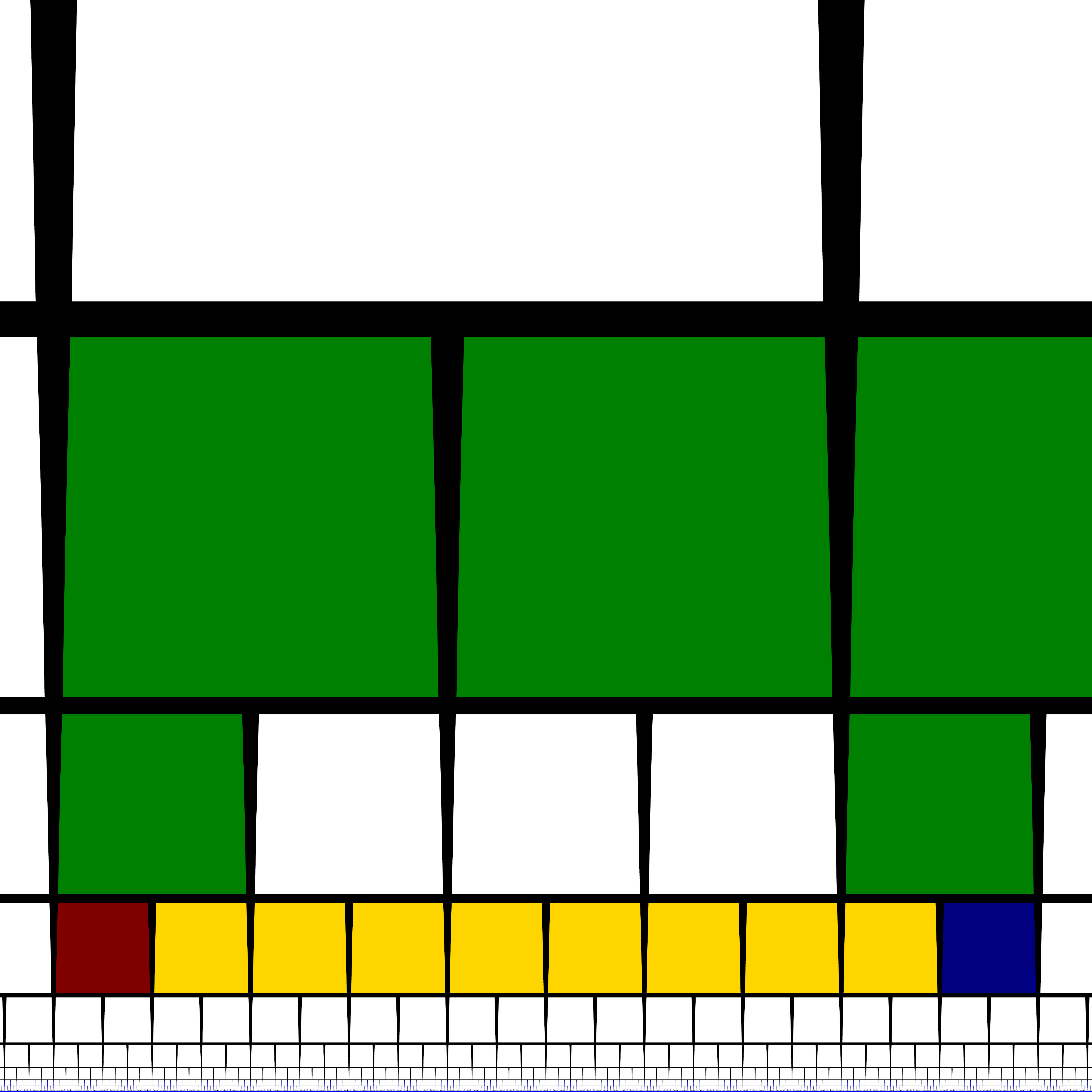}
\includegraphics[width=0.22\linewidth]{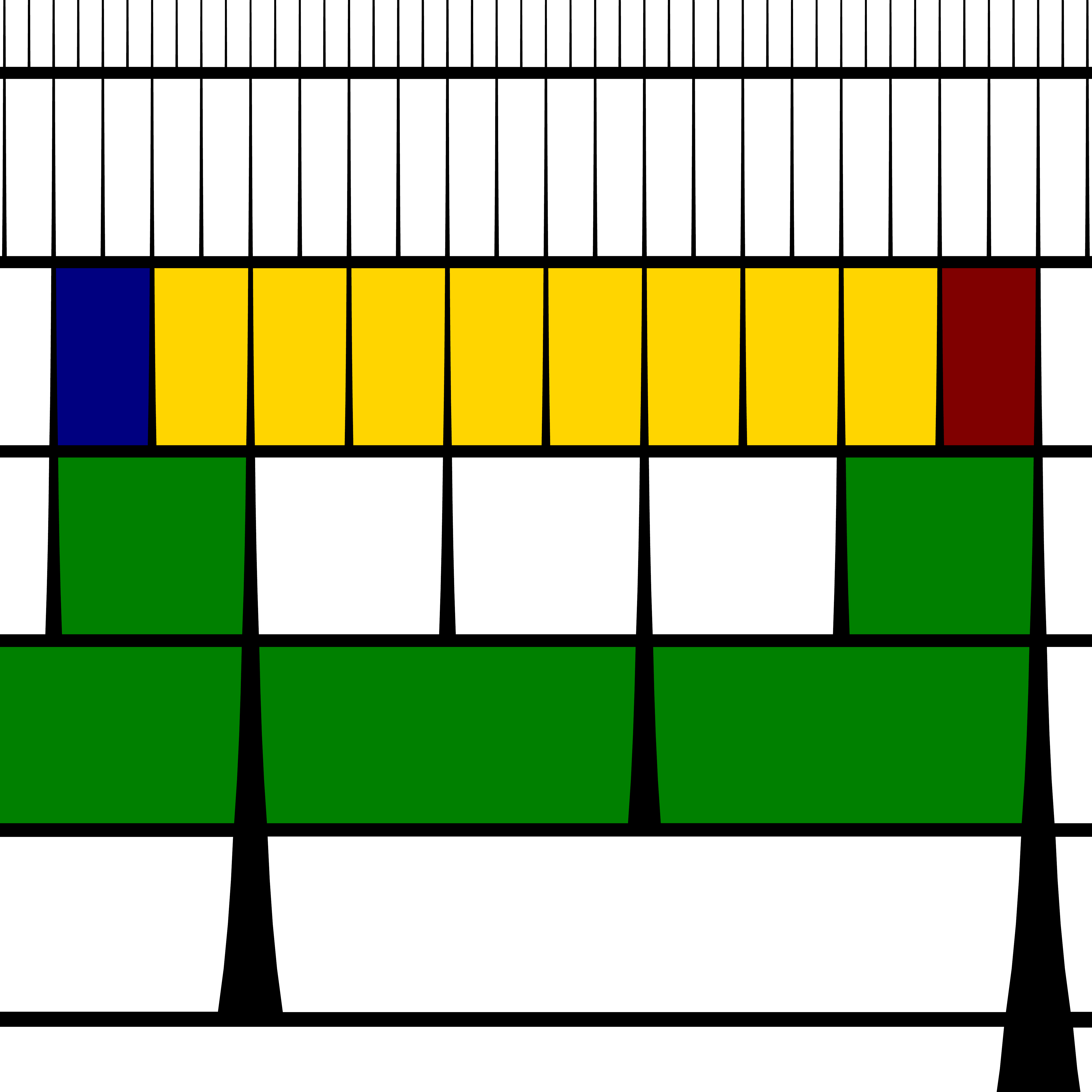}
\caption{\label{figlet} Binary tiling of $\bbH^2$, in four projections. From left to right: Poincar\'e disk, Beltrami-Klein disk, Poincar\'e half-plane, Horospherical.}
\end{figure*}

Non-Euclidean geometries may be perceived as unnatural and confusing to navigate. For two-dimensional
geometries, there are numerous projections (models). However, one may subject to the trade-off
between the comprehensibility (taming strangeness) and the ease of finding the shortest straight line (crucial for convenient navigation). To illustrate this, compare several projections of the same scene in two-dimensional hyperbolic geometry in Fig.~\autoref{figlet}. Although half-plane and horocyclic projections
seem easier to understand, they misperceive the straight yellow line as the shortest route between the red and blue cells instead of the shorter green line. On the contrary, geodesic-based, azimuthal projections make finding the shortest path straightforward.

For visualization of three-dimensional geometries, the aforementioned trade-off is mitigated with first-person perspective. We put structures in a given geometry, and we render
how a person inside the geometry would view those structures. We assume that light rays always travel
along the shortest routes in our space (geodesics). Such an approach aces in applications where finding the shortest path between two points is critical. Also, positioning of the observer inside the geometry should 
help them familiarize with a new environment.

In comparison to two-dimensional non-Euclidean geometries, non-isotropic three-dimensional geometries, {\bf Solv}, {\bf Nil}, and twisted $\bbH^2\times\bbR$, are even more demanding to comprehend. Weeks \cite{weeks2001shape} describes the {\bf Solv} geometry as
\emph{``This is the real weirdo. [...] I don't know any good intrinsic way to understand it.''}. Therefore, efficient visualization becomes a fundamental tool for gaining intuition about those geometries.
From the programmer's point of view, there are two major challenges in visualizing non-isotropic geometries. (1) the geodesics in these geometries are not
necessarily given by simple formulae (especially in {\bf Solv}), (2) for given points $a$ and $b$, there can be multiple geodesics from $a$ to $b$. As a result, while there are implementations of real-time first-person view for
Euclidean,
spherical, hyperbolic spaces \cite{weeksrealh,hyperbolicvr}, and for product spaces
\cite{weeksreal}, real-time visualizations of geometries like {\bf Solv}, {\bf Nil} or twisted $\bbH^2\times\bbR$ were absent until recently.

This paper presents novel methods of real-time native geodesic rendering of first-person perspective in
non-isotropic three-dimensional geometries.
Our solution has major advantages over
other propositions. First, our proposed method outreaches pure visualization, allowing for convenient hands-on activities, for example in video games, education or art, as well 
as applications in machine learning and physics simulations. 
Second, our primitive-based method is better suited for Virtual Reality. 
Our implementation is also the only one suitable for working with large-scale scenarios; other implementations would not be suitable
because of numerical issues inherent to negatively curved spaces.

The paper proceeds as follows. Section 2 introduces definitions necessary for understanding non-isotropic geometries, illustrated by the examples from simpler cases of two-dimensional and isotropic geometries. In Section 3 we
present our method. We start with describing technical components of our primitive-based method;
next we discuss the details for rendering particular non-isotropic geometries. In Section 4, we evaluate
our results, providing analysis of errors and comparision with competitive approach. Section 5 discusses
our contribution and presents possible areas of application, and Section 6 concludes.


\section{Theoretical background}

\subsection{Riemannian manifolds and isotropic geometries}

We will use a simplified definition of a Riemannian manifold. While 
less general than the commonly used definition, our definition is convenient for computations and satisfies our needs.
Intuitively, a Riemannian manifold is a $n$-dimensional subset of $\bbR^m$ which locally behaves like an $n$-dimensional Euclidean space.
Let $\bbR_{m,m}$ be the set of bilinear functions from $\bbR^m \times \bbR^m$ to $\bbR$.

\begin{definition}
An $n$-dimensional \emph{(Riemannian) manifold} is $M = (A,g)$, where: 
\begin{itemize}
\item $A \subseteq \bbR^m$,
\item $g: A \ra \bbR_{m,m}$ ($g(x)$ is a bilinear function),
\item for every $x \in A$ there is a open neighborhood $U\subseteq \bbR^m$ of $x$
and a differentiable bijection $f: V \ra U \cap A$, where $V$ is a open neighborhood of 0 in $\bbR^n$, such that 
for every vector $0 \neq v \in \bbR^n$, $g(x)(Df(x), Df(x))>0$.
\end{itemize}
\end{definition}

The bilinear function $g(x)$ is known as the metric tensor and is used to measure the length of curves. Let $\gamma:[t_1,t_2] \ra A$ be a curve (i.e, a continuous
differentiable function). We define the length of $\gamma$
\longonly{ using the following formula: ($\dot\gamma$ is the derivative of $\gamma$) \[ l_M(\gamma) = \int_{x=0}^t \sqrt{g(\dot\gamma(x), \dot\gamma(x)))} dx. \]}%
\shortonly{ using the formula $l_M(\gamma) = \int_{x=0}^t \sqrt{g(\dot\gamma(x), \dot\gamma(x)))} dx$ where $\dot\gamma$ is the derivative of $\gamma$.}

\begin{definition}
We say that manifolds $M_1=(A_1,g_1)$ and $M_2=(A_2,g_2)$ are {\bf isometric} if and only if there is a bijection $f: A_1 \ra A_2$ such
that for every curve $\gamma$, $l_{M_1}(\gamma) = l_{M_2}(f(\gamma))$.
\end{definition}
When $M_1$ and $M_2$ are isometric, we consider them to be different models of the same abstract manifold.

\begin{definition}
A \emph{geodesic} is a curve $\gamma$ that is locally shortest and constant speed. For every $t$, there is an interval $(t_1, t_2) \ni t$
such that for $t_1<u_1<u_2<t_2$, $\gamma$ restricted to $[u_1,u_2]$ is the shortest curve from $\gamma(u_1)$ to
$\gamma(u_2)$. Moreover, $g(\dot\gamma(x), \dot\gamma(x))$ is a constant.
\end{definition}

\begin{definition}
A \emph{geometry} is a manifold that is complete, simply connected, and locally homogeneous.
A manifold $(A,g)$ is \emph{simply connected} if and only if for every two points $x,y \in A$, there exists a curve
from $x$ to $y$, and every two curves from $x$ to $y$ are homotopic, i.e., one can be continuously deformed into the
other; \emph{locally homogeneous} if and only if for every two points $x,y \in A$, there exist 
open neighborhoods $X \ni x, Y \ni y$ such that $(X,g)$ and $(Y,g)$ are isometric;
\emph{complete} if and only if every geodesic $\gamma: [t_1, t_2] \ra A$ can be extended to $\gamma: \bbR \ra A$.
\end{definition}

In a locally homogeneous manifold, every point locally looks the same. A manifold is called \emph{isotropic} if
additionally it looks the same in every direction. The following isotropic geometries exist for every dimension $n \geq 2$:

\begin{description} 
\item[Euclidean geometry] $\bbE^n$ given by $A = \bbR^n$ and $g(x)(v,w) = v \cdot w$, where $\cdot$ is the inner product.

\item[Spherical geometry] $\bbS^n$ given by $A = \{v \in \bbR^{n+1} : v \cdot v = 1\}$ and $g(x)(v,w) = v \cdot w$. 
This is the surface of a sphere in $n+1$-dimensional space.

\item[Hyperbolic geometry] $\bbH^n$ given by $A = \{v \in \bbR^{n+1} : v \cdot v = -1, v_{n+1} > 0\}$ and $g(x)(v,w) = v \cdot w$,
where $\cdot$ is the Minkowski inner product: $(x_1, \ldots, x_{n+1}) \cdot (y_1, \ldots, y_{n+1}) = 
x_1y_1 + x_2y_2 + \ldots + x_ny_n - x_{n+1}y_{n+1}$.
\end{description}

We have described the hyperbolic geometry in the Minkowski hyperboloid model. To explain the $\Sol$ geometry,
we will need also other models \cite{cannon}:

\begin{description}
\item[Beltrami-Klein model] where $A_1 = \{v \in \bbR^n : v \cdot v \leq 1\}$, is obtained from the Minkowski hyperboloid
model via the map $f(h) = (h_1/h_{n+1}, \ldots, h_n/h_{n+1})$. The metric tensor $g$ is defined in the unique way that
yields an isometry.

\item[Poincar\'e ball model] where $A_2 = \{v \in \bbR^n : v \cdot v \leq 1\}$, is obtained from the Minkowski hyperboloid
model via the map $f(h) = (h_1/(1+h_{n+1}), \ldots, h_n/(1+h_{n+1}))$. The metric tensor $g$ is defined in the unique way that
yields an isometry.

\item[Half-space model] where $A_3 = \{v \in \bbR^n : v_n > 0\}$, is obtained from the Poincar\'e ball model via
inversion in a circle centered at $(0,\ldots,0,-1)$. The metric tensor $g$ is defined in the unique way that
yields an isometry; we get $g(x)(v,w) = x_{n}^2 (v \cdot w)$.

\item[Horospherical model] where $A_4 = \bbR^n$, is obtained from the half-space model via
$f(x_1,\ldots,x_n) = (x_1,\ldots,x_{n-1},\log x_n)$. The metric tensor $g$ is defined in the unique way that
yields an isometry; we get $g(x)(v,w) = e^{2x_n}v_1w_1 + e^{2x_n}v_2w_2 + \ldots + e^{2x_n}v_{n-1}w_{n-1} + v_nw_n$.
\end{description}

\longonly{For two manifolds $\bbA = (A,g_A)$ and $\bbB = (B,g_B)$, their product manifold $\bbA\times\bbB$ is 
$(A \times B, g)$, where, for every $a_1,a_2 \in A$, $b_1,b_2 \in B$, $g((a_1, b_1),  (a_2, b_2)) =
g_A(a_1,a_2) + g_B(b_1,b_2)$. Note that $\bbE^n \times \bbE^m = \bbE^{n+m}$.}

\subsection{Tangent spaces, geodesics, and parallel transport}\label{tangentspace}

The tangent space $T_a(A)$ is the set of vectors $v\in\bbR^m$ such that there exists a curve $\gamma: \bbR \ra A$ such that $\gamma(0)=a$
and $\dot\gamma(0) = v$. Let $a \in A$ and $v \in T_a(A)$, the exponential map $\exp_a(v)$ is $\gamma(1)$, where $\gamma$ is the unique
geodesic such that $\gamma(0) = a$ and $\frac{d}{dt}(\gamma(0)) = v$. Intuitively, $\exp_a(v)$ tells us where we end up if we start
in the point $a$ and follow the geodesic in the direction and distance given by $v$. The inverse of $\exp_h$ is the 
inverse exponential map $\log_a: A \ra \bbR^m$. This may be a multivalued function (similar to $\log(b)$ in the complex plane).
 The inverse exponential map $\log_a(b)$ tells us which direction 
(and distance) we should go in order to reach $b$ from $a$.

Let $\gamma$ be a curve such that $\gamma(t_0)=a$ and $\gamma(t_1)=b$. Parallel transport lets us move tangent vectors $v \in T_a(A)$
to $\in T_b(A)$ along the curve $\gamma$ in a natural way. Contrary to Euclidean space, the resulting $w \in T_b(A)$ may depend on the choice of $\gamma$.
For example, in $\bbH^2$ and $\bbS^2$, the sum of internal angles in a triangle is $180^\circ+\epsilon$ where $\epsilon<0$ in $\bbH^2$
and $\epsilon>0$ in $\bbS^2$. As a consequence, if we walk on a loop $\gamma$ which cycles around such a triangle, we need to turn by
$360^\circ-\epsilon$ angles in total. Thus, the vector $v \in T_a(A)$ will be transported to $v' \in T_a(A)$, where $v'$ is $v$ rotated by
angle $\epsilon$. In general, the rotation equals the area enclosed in $\gamma$ times the curvature (or integral of the curvature 
for non-homogeneous manifolds).

To compute the exponential function and parallel transport in non-isotropic manifolds we will be using the Christoffel symbols. Assume $n=m$ (otherwise
use another model that has this property). Let $g^{ij}$ be the matrix of coefficients of $g$, $g_{ij}$ be the
inverse of this matrix, and $\partial_i = \frac{\partial}{\partial x_i}$. The Christoffel symbols are given by 
\begin{equation} \label{christoffel}
\Gamma^k_{ij} = {\frac{1}{2}} \sum_m g^{km} (\delta_i g_{mj} + \delta_j g_{im} - \delta_m g_{ij}).
\end{equation}
Parallel transport is given by the following system of differential equations: $v(t_0)=v$ and
\begin{equation} \label{ptrans}
\dot v^k = -\sum_i \sum_j v^i \dot\gamma^j \Gamma^k_{ij}.
\end{equation} 
The curve $\gamma$ is
a geodesic if and only if the above hold for $\dot\gamma$, i.e., 
\begin{equation} \label{geoeq}
\ddot \gamma^k = -\sum_i \sum_j \dot\gamma^i \dot\gamma^j \Gamma^k_{ij}.
\end{equation}

\section{Technical details of our method}

\subsection{Homogeneous coordinates} \label{homoco}
In computer graphics, we commonly represent $n$-dimensional Euclidean space using homogeneous coordinates $(x_1,\ldots,x_n,x_{n+1})$,
where $x_{n+1}$ = 1. This lets one represent both translations and rotations as matrix multiplications. The same property also holds in the
spherical and Minkowski hyperboloid coordinates. Rotations around the homogeneous origin $h_0 = (0...0,1)$ are described
by the same matrices in all 
three geometries. Translation by $x$ along the first axis does not change the coordinates except $x_1$ and $x_{n+1}$, while 
$x_1$ and $x_{n+1}$ are affected in the following way: $\left( \begin{array}{c} x'_1 \\ x'_n+1 \end{array} \right) = 
M \left( \begin{array}{c} x_1 \\ x_n+1 \end{array} \right)$, where the matrix $M$ is of form:
\begin{itemize}
\item $\left( \begin{array}{cc} 1 & x \\ 0 & 1 \end{array} \right)$
in Euclidean geometry,
\item $\left( \begin{array}{cc} \cos x & \sin x \\ -\sin x & \cos x \end{array} \right)$ in spherical geometry,
\item $\left( \begin{array}{cc} \cosh x & \sinh x \\ \sinh x & \cosh x \end{array} \right)$ in hyperbolic geometry.
\end{itemize}

These formulas make visualizations of isotropic geometries, including the camera movement, a straightforward generalization of the Euclidean methods \cite{gunnvis}.%
\longonly{
According to our experience,
beginners in hyperbolic rendering tend to use the Poincar\'e model, since that model is commonly used in courses,
while the Minkowski hyperboloid model is ignored. Later, they learn about the Minkowski hyperboloid model and find out that it is easier
to understand by analogy to spherical geometry and also due to its much better numerical properties.
}

In Euclidean 3D graphics it is said that it is easier to move and rotate the world than to move and rotate the camera. The same stays true in other isotropic geometries.
In non-isotropic geometries we can still move our point of vision by moving the world, and to make this work we will also use
coordinate systems where translations are represented by matrix multiplications. However, we can no longer rotate the world (rotations are no longer isometries), so we will represent the camera orientation as a triple of vectors (top, right, front directions,
denoted $d_1, d_2, d_3 \in T_c(A)$ where $c$ is the camera position), or equivalently a view matrix $V$ such that $Ve_i = d_i$ and
$Ve_4 = e_4$, where $e_i$ is the $i$-th unit vector.
To find out the screen coordinates of an object located at $x$, we apply the perspective projection to $(TV)^{-1} \log_{h_0}(Tx)$, where $T$ is
the translation matrix that moves the current camera position $c$ to the homogeneous origin $h_0$. The camera can be rotated in the standard
way. To move the camera $d$ units forward, we compute the geodesic $\gamma$ such that $\gamma(0)=c$ and $\dot\gamma(0)=d_3$. The new camera
position will be $\gamma(d)$, and the new camera orientation $(d_i)$ is computed by parallel transport of respective $(d_i)$ along the
geodesic $\gamma$. This keeps the front vector $d_3$ always point forward as we traverse our geodesic; for vectors $d_1$ and $d_2$,
using parallel transport ensures that the camera is not weirdly rotated as we travel.

While camera movement is an important component of immersive visualization, it also turned out to be very helpful as a tool to verify
the correctness of our renderers. If both the camera movement and the renderer are working correctly, the object seen in the exact center of the screen
should remain in the exact center as we move the camera forward. If this is not the case, at least one component is not working correctly.

\subsection{Tessellations}
A {\it tessellation} of a manifold is its tiling using a compact shape (called \emph{tile} or \emph{cell})
with no overlaps or gaps. Three-dimensional tessellations are called honeycombs.
In general, tessellations may use multiple shapes; the tessellations in
our paper will always use just one. For example, the Euclidean plane can be tessellated with squares or with regular hexagons.

In non-Euclidean visualizations, tessellations serve two important goals. First, 
they can be used as landmarks or milestones to help the observer 
navigate and measure the space. Second, due to the exponential expansion of $\bbH^d$, (twisted) $\bbH^2\times\bbR$ and $\Sol$, 
representations based on floating point coordinates are very prone to precision errors. Tessellations can be represented
in a discrete way, providing a way to avoid this issue.

We will describe the tesselations used for particular non-isotropic geometries in the respectible sections of this paper. Here, we will explain
the general idea behind generating tesselations for the needs of our implementation using
the simplest tessellation of the hyperbolic plane, i.e., the \emph{binary tiling} \cite{boroczky}. We will use the binary tiling later
in our tessellation of the $\Sol$ geometry.

In the horospherical model, the shape is given as $S = [0,1] \times [0,\log 2]$.
By translating S with the isometry $f_k(x,y) = (x+k,y)$ for $n \in \bbZ$
we tessellate the horoannulus $S = \bbR \times [0, \log 2]$; by translating in two dimensions using the isometry $f_{k,l}(x,y) = (2^{-l}x+k, y+l \log 2)$,
we tessellate the whole hyperbolic plane. The binary tiling has a structure similar to that of the infinite binary tree (Fig.~\autoref{figlet}).

Every tessellation used in our implementation can be seen as a combinatorial graph that can be generated lazily. 
In the case of the binary tiling,
every tile has pointers to its five
neighbors (left, right, up-left, up-right, down); these pointers are initially null pointers and point to specific tiles
once the relevant tile is known. We start with a single root tile and generate new tiles as required, using simple rules:

\begin{itemize}
\item if we are asked about an unknown up-left or up-right neighbor of $X$, we create a new tile $Y$. We connect $X$ to $Y$
(setting $Y$ as the up neighbor of $X$, and $X$ as the down neighbor of $Y$).
\item the left neighbor of $X$ is the up-left neighbor of the down neighbor of $X$ (if $X$ is the up-right neighbor of its
down neighbor), or the up-right neighbor of the left neighbor of the down neighbor of $X$  (if $X$ is the up-left neighbor).
The right neighbor can be found similarly.
\item if we are asked about an unknown down neighbor of $X$, we create a new tile, and we arbitrarily assign $X$ as one of
the upper neighbors of $Y$.
\end{itemize}

\begin{figure}[ht]
\begin{center}
\includegraphics[width=.5\linewidth]{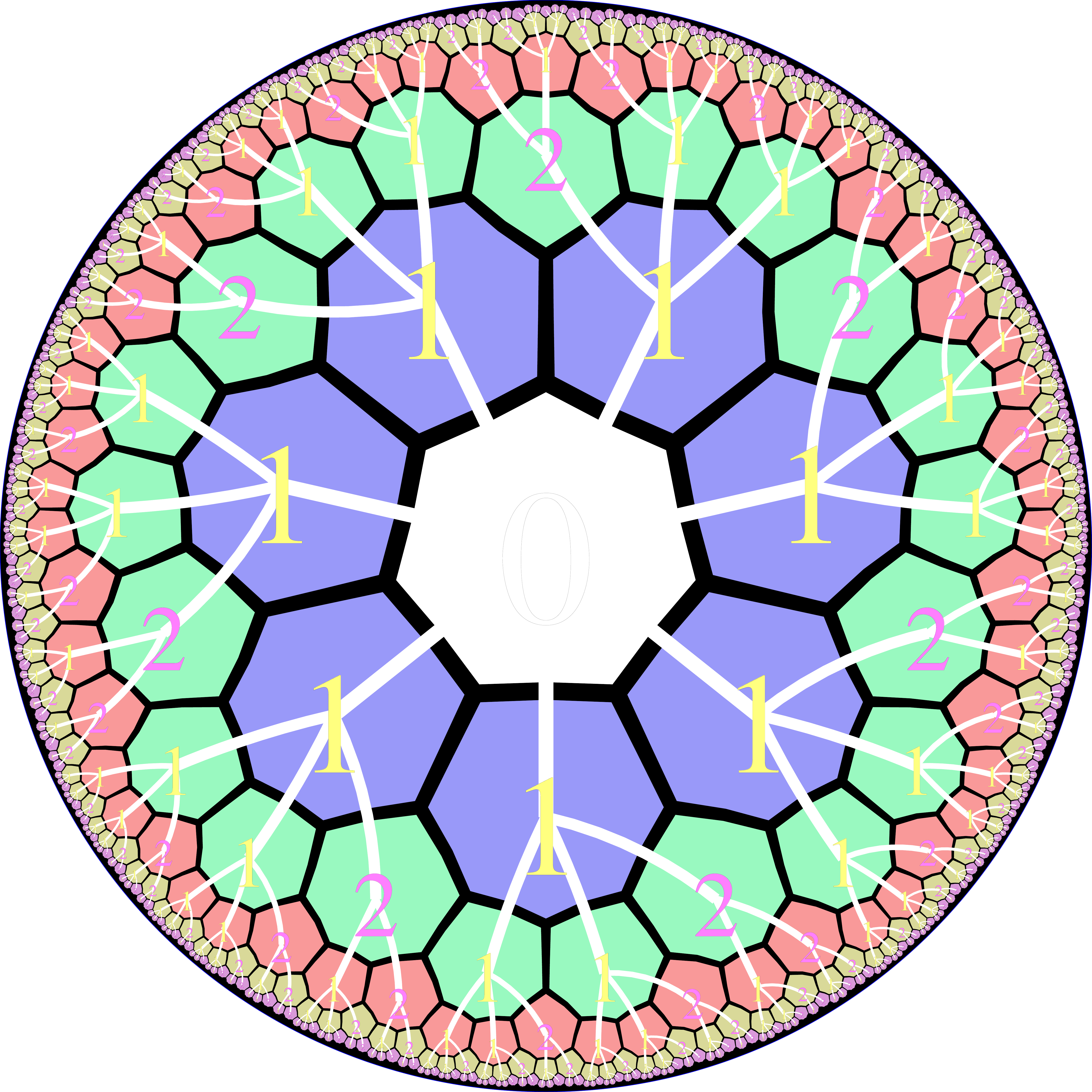}
\end{center}
\caption{Tree structure used to generate \{7,3\}.\label{treegen}}
\end{figure}

Similar, but somewhat more complicated rules can be also used to generate regular tilings of the hyperbolic plane.
We use the tree shown in Figure \ref{treegen} to generate the tessellation $\{7,3\}$, whose every cell is a regular
heptagon, with three meeting in every vertex. The central tile has type 0 and seven children, every other tile
has type $t \in \{1,2\}$, and $4-t$ children. There are simple rules to tell the types of children of the given
vertex, as well to find the cells connected; see e.g., \cite{margenstern_heptagrid,trigridold} for more details.
Representing points in $\bbH^2$ with a pointer to the tile they are in, and Minkowski coordinates relative
to that tile, lets us avoid numerical issues that appear when representing faraway points using only model coordinates.

\subsection{Non-Isotropic geometries}

\subsubsection{Product geometries}\label{sec:prod}

\begin{figure}[ht]
\begin{center}
\includegraphics[width=.5\linewidth]{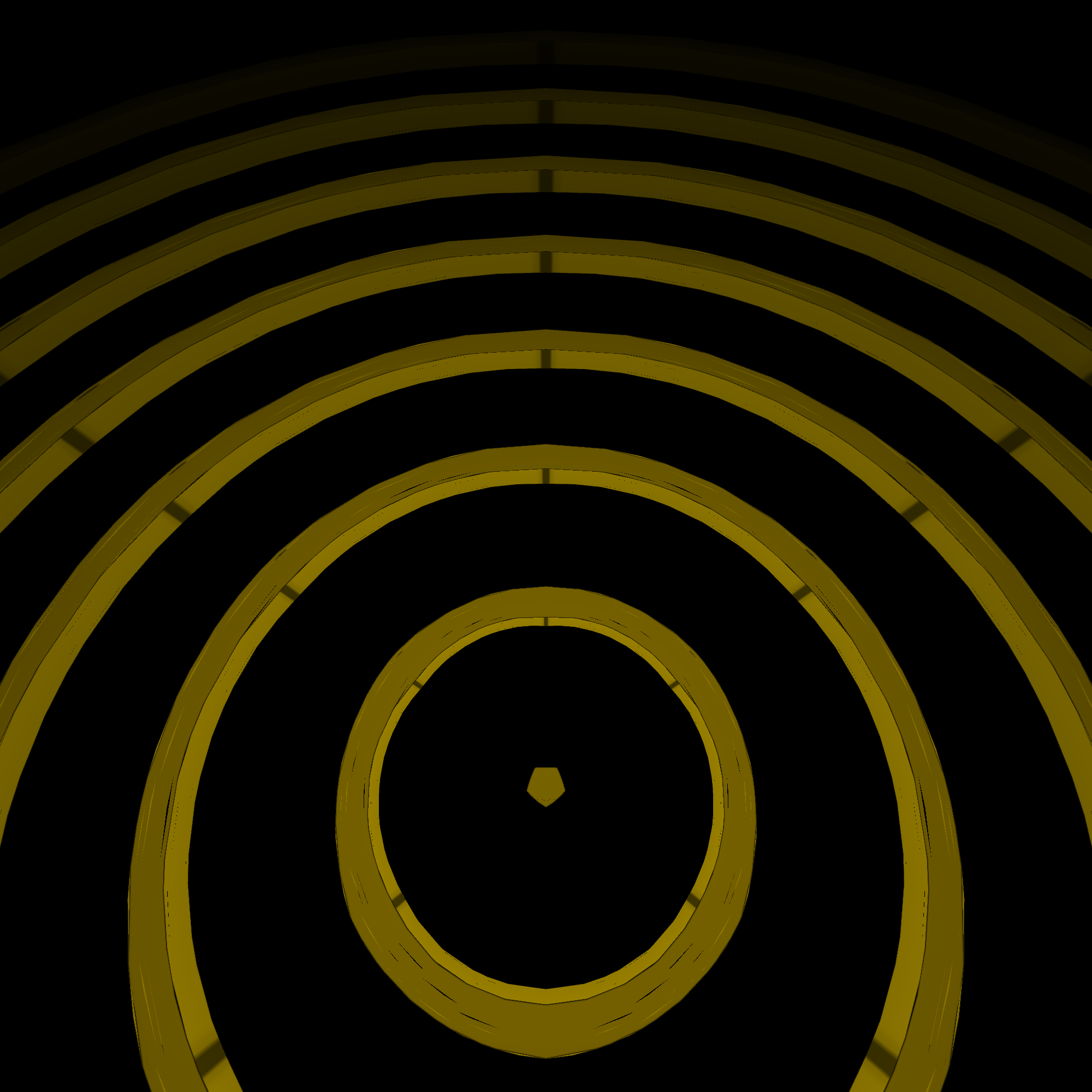}
\end{center}
\caption{A single prism in $\bbS^2 \times \bbR$ geometry. \label{figse}}
\end{figure}

For the product geometry $\bbH^2\times\bbR$ (and similarly $\bbS^2\times\bbR$), taking the product of homogeneous representations of $\bbH^2$ and $\bbR$
yields a five-dimensional homogeneous representation, which is difficult to work with in OpenGL. However, there is also a three-dimensional
coordinate system which has the desired property: for $h \in \bbH^2$ and $x \in \bbR$, we represent $(h,e)$ as $h \cdot \exp x \in \bbR^3$.

In both $\bbH^2 \times \bbR$ and $\bbS^2 \times \bbR$, a honeycomb structure is obtained as a product of a tessellation in the underlying two-dimensional
geometry, and a slicing of the $\bbR$ coordinate (into levels of the same height). Therefore, every cell in our construction is identified by two
generalized coordinates: cell in the two-dimensional tessellation, and the z-level.

The implementation of $\bbH^2 \times \bbR$ geometry does not pose any major new technical nor mathematical challenges.
Computing $\exp_{(v,z)}$ and $\log_{(v,z)}$ is straightforward. $\bbS^2 \times \bbR$ is more challenging. Figure \autoref{figse} 
presents a single pentagonal prism in the $\bbS^2 \times \bbR$ geometry. The prism is directly below the camera. The prism
is visible as a series of concentric rings. This is because $\exp_{v,0}(2\pi k \cos(\alpha), 2\pi k \sin(\alpha), z) = (v,z)$ for every $k \in \bbZ$ and $\alpha \in \bbR$.
The light ray goes around the sphere $k$ times before hitting our prism. Triangles close to the points directly below or above the camera, or its antipodal point,
may appear like crescent shapes or rings. To render these shapes correctly, such triangles are subdivided. This process is very technical, so we omit the description in this paper.

\subsubsection{{\bf Solv} geometry}

\begin{figure}[ht]
\begin{center}
  \includegraphics[width=0.5\linewidth]{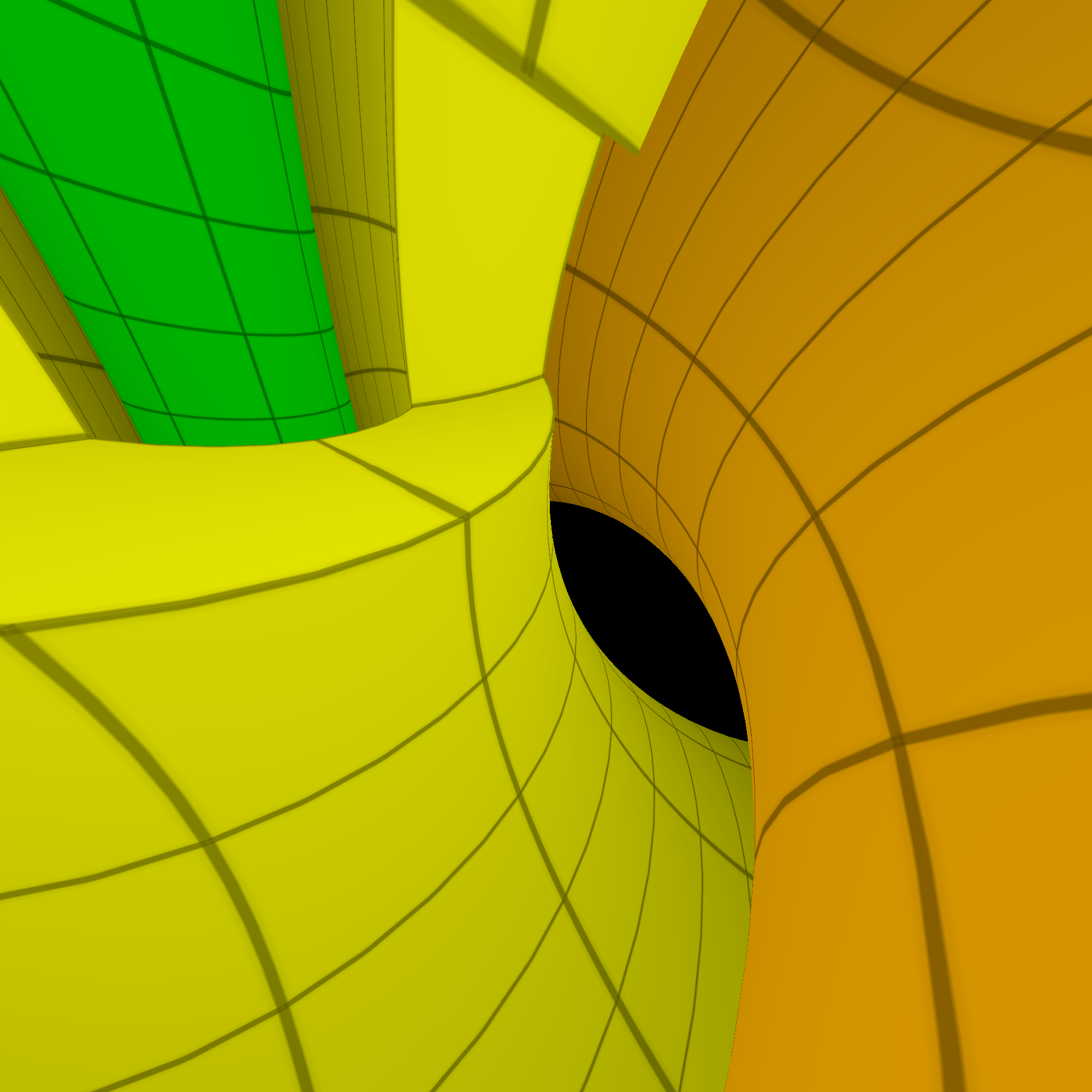}
\end{center}
\caption{A scene in Solv geometry.\label{solvscene}}
\end{figure}

Among the geometries we visualize, the $\Sol$ geometry is the most challenging. 
Our implementation is based on the paper by Bölcskei and Szilágyi \cite{solvgeodesics}. This geometry is $(\bbR^3, g)$, where
$g^{11}(x,y,z) = \exp z$, $g^{22}(x,y,z) = \exp {-z}$, $g^{33}(x,y,z) = 1$, and $g^{ij}=0$ for $i \neq j$.
The isometry taking $(x,y,z)$ to $(0,0,0)$ is given by $m(x',y',z') = (x-e^zx', y-e^{-z}y', z-z')$; as explained in Section \ref{homoco}
we add the fourth homogeneous coordinate, always equal to 1, so that this isometry can be represented as matrix multiplication. (We will ignore this fourth
coordinate below.)
Note that the plane $y=0$ is the hyperbolic plane in the horospherical model, and the plane $x=0$ is also the hyperbolic plane
in the horospherical model, but where the coordinate $z$ is reversed. The plane $z=0$ is Euclidean.

To understand the geodesics in {\bf Solv}, consider special cases. What is the shortest curve from $(0,0,0)$ to $(M,0,0)$ (where $M$ is large)?
Both points belong to the hyperbolic plane $y=0$, thus the geodesic will act just like in this hyperbolic plane (we can easily see
that we cannot obtain a shorter curve by changing $y$). We already know how geodesics work in the horospherical coordinates model:
the obvious curve $\gamma(t) = (t,0,0)$ is not the shortest (and thus not a geodesic),
because its length is $M$, and we get a shorter curve by moving first to $(0,0,-\log M)$, then to $(M,0,-\log M)$, then to $(M,0,0)$.
The total length of this curve is $\log M + 1 + \log M = 2\log M+1$; the actual geodesic is obtained from the $\bbH^2$ geodesic by adding
the extra coordinate $y=0$. Its construction is similar to that of the polyline constructed above; in particular, its length 
is also $\Theta(\log M)$.

Similarly we can find the shortest curve from $(0,0,0)$ to $(0,M,0)$; however, since in the hyperbolic plane $x=0$, the coordinate $z$
is reversed, our polyline will first move to $(0,0,\log M)$. We can find the geodesics from $(0,0,0)$ to any point $(x,0,z)$ or $(0,y,z)$
by adding a zero coordinate to the respective geodesic in $\bbH^2$.

The situation is more difficult for points $(x,y,z)$ where $x,y \neq 0$. In particular, let us try to find the shortest curve from $(0,0,0)$
to $(M,M,0)$. While the curve $\gamma(t) = (t,t,0)$ is a geodesic of length $M\sqrt{2}$, it is not the globally shortest one.
There is a polyline of length $4 \log M+2$, which goes through the following points: $(0,0,0) - (0,0,-\log M) - (M,0,-\log M)
- (M,0,\log M) - (M,M,\log M) - (M,M,0)$. There is also another polyline of the same length, going through the points: $(0,0,0) - 
(0,0,\log M) - (0,M,\log M) - (0,M,-\log M) - (M,M,-\log M) - (M,M,0)$. It can be seen that the actual geodesic will again be of 
similar nature to one of these polylines: we have to temporarily increase the $z$ coordinate in order to traverse the large difference in
$y$ coordinate quickly, and also to temporarily decrease the $z$ coordinate to traverse the large difference in $x$. However, these two
movements can be done in any order, yielding two distinct geodesics (by symmetry, of the same length), one of which starts almost precisely
upwards (for large values of $M$), and the other starts almost precisely downwards. For points $(M_1,M_2,0)$ where $|M_1| \neq |M_2|$,
one of these geodesics will be shorter.

To determine the actual geodesics we need to solve the geodesics equations (\ref{geoeq}). This has been done by  Bölcskei and Szilágyi \cite{solvgeodesics};
however, the result obtained is in terms of integrals of elliptic functions, and it is not clear how to compute it efficiently. Therefore,
we determine the exponential function $\exp_0$ by solving the geodesic equation (\ref{geoeq}) numerically. We use the Runge-Kutta RK4 method 
with 100 steps.

\begin{figure}[ht]
\begin{center}
\includegraphics[width=0.7\linewidth]{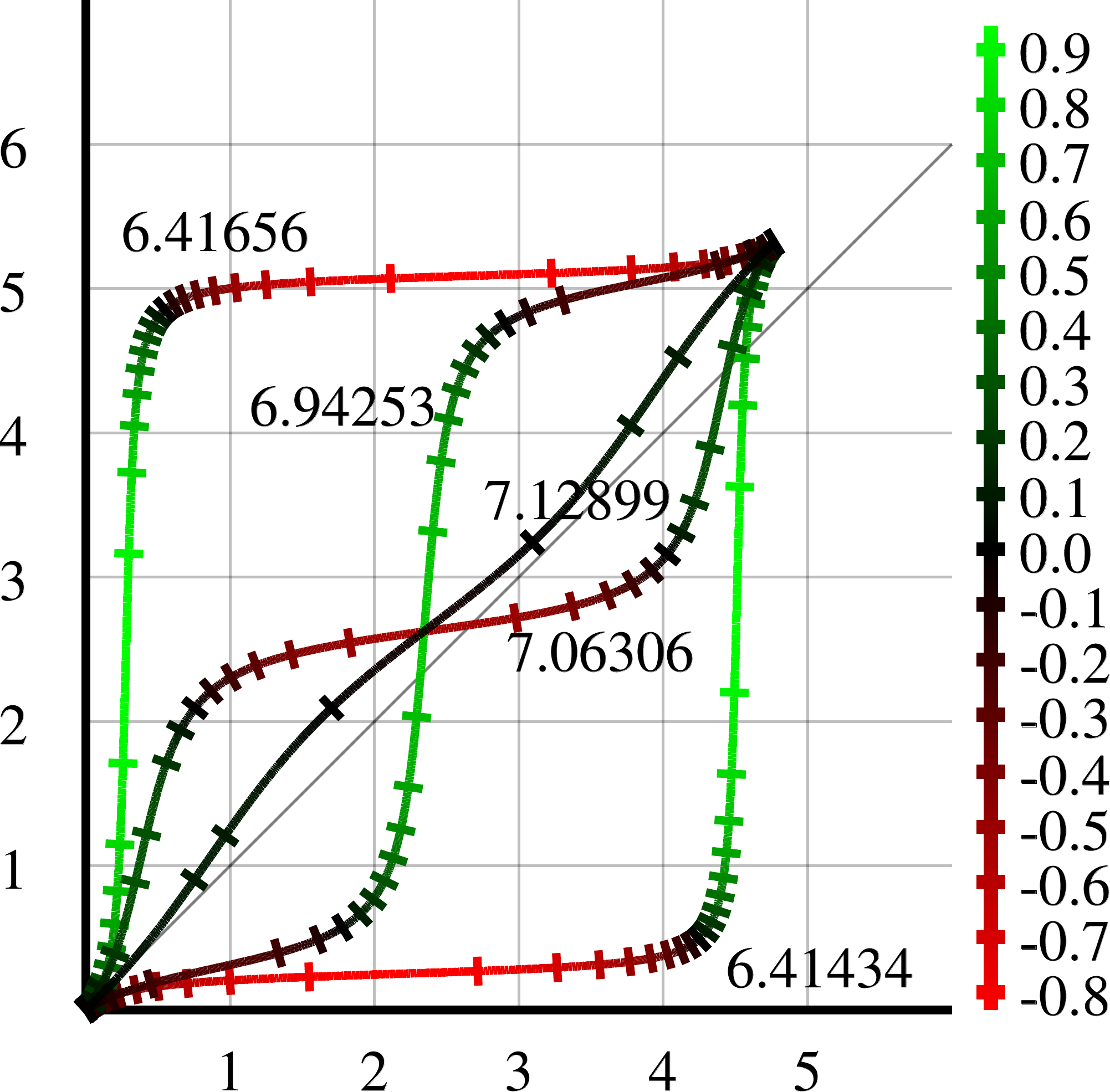}
\end{center}
\caption{Geodesics in {\bf Solv}. Color is used to visualize the third dimension. \label{solvgeo}}
\end{figure}

Figure \autoref{solvgeo} shows the graphs of several geodesics from (0,0,0) to $(x,y,z)$ and their lengths; the geodesics are three-dimensional,
color is used to visualize the third dimension.
Other than the geodesics strictly embedded in hyperbolic planes ($x(t)=C$ or $y(t)=C$),
$x(t)$ and $y(t)$ are strictly monotonous with time, while $z(t)$ is periodic: it increases to the top level $z_1$,
afterwards it decreases to $z_2$, then back to $z(t)$. The derivatives $\dot x(t)$ and $\dot y(t)$ are functions of $z(t)$ ($|\dot x(t)|$
is bigger when $z(t)$ is small and smaller when $z(t)$ is big).

To render {\bf Solv} we need to find $\log_M(a)$. We use an iterative method similar to the Newton method. In the $n$-th iteration,
we compute $\exp_0(t_n)$ and $\exp_0(t_n+\epsilon e_i)$ for $i=1,2,3$ (we use $\epsilon = 10^{-6}$). This lets us find an affine function $f$
which agrees with $\exp_0$ in the four testing points; the next $t_{n+1}$ will be such that $f(t_{n+1}) = a$. For the interesting values of $a$,
we have verified experimentally that this method quickly and successfully finds $t$ such that $\exp_0(t) = a$, if we start with $t_0 = 0$
and limit the step size to 0.1.

While the method above always finds a geodesic, it might not find the shortest geodesic. For example, for $a = (x, x, 0)$ it will find $t=a$
which is of length $x\sqrt{2}$, while the shortest geodesic is of length $\Theta(\log x)$. For our applications,
finding the shortest geodesic is the most important. The problem is present for $a = (x,y,z)$ where both $|x|$ and $|y|$ are large. 
In this case, the shortest geodesic has a structure similar to the shortest paths described earlier in this section: we move through
to a point $b$ close to either $p_1=(x,0,\frac{z}{2})$ or $p_2=(0,y,\frac{z}{2})$. To find the actual shortest geodesic, we need to find the
point $b$, of form $b = p_i + (x,-x,z)$. Since this is enough to find an arbitrary point on the geodesic, we find one in the intersection of the
geodesic with the hyperplane $\{p_i+(x,-x,z):x,z \in \bbR\}$. We set $b_0 = p_i$ for $i=1,2$, and then iteratively minimize the sum of geodesic
distance from 0 to $b_n$ and from $b$ to $a$; this can be done by computing $f(b) = |\log_0(b)| + |\log_0(I_b a)|$ where $I_b$ is an isometry that
takes $b$ to 0. In both cases we compute $\log_b$ for a point that has only one coordinate distant from 0, and for such points the Newton method
described above works. We minimize $f(b)$ by approximating first-order and second-order derivatives of $f$, and finding the minimum of the
obtained quadratic function. Once $b$ is found, $\log_0 a$ can be computed as
$t' = \frac{|\log_0(b)|+|\log_0(I_ba)|}{|\log_0(b)|}\log_0 b$;
to combat the precision issues, we find the actual $t$ by the Newton method, starting the iteration from $t'$.

The methods described above are too computationally expensive for real-time visualization. We combat this by constructing a $D \times D \times D$
table of precomputed values, and then use interpolation. We take $D=64$. Such interpolation can be performed efficiently on GPU hardware (in GLSL, the table is
loaded as a texture). Since $x$, $y$, $z$ are unbounded, we will actually precompute a function $g$ such that $\log_0(x,y,z) = k^{-1}(g(i_x(x), i_y(y), i_z(z)))$,
where $i_x(x), i_y(y), i_z(z) \in [0,1]$. We only consider $x,y,z \geq 0$ (we can use symmetry to compute $\log_0$ for negative
arguments). For $i_x(x)$ we map the point $(x,0)$ in the horocyclic coordinates to the
Poincar\'e disk model; in the Poincar\'e disk model, the horocycle is mapped to a circle (see \autoref{figlet}), and $i_x(x)$ is the angular coordinate on that (semi-)circle,
scaled to $[0,1]$. Function $i_y$ works in the same way. 
Our $i_z(z)$ is the Poincar\'e disk coordinate of the horocyclic point $(0,z)$. 
The function $k$ maps points in $\bbR^3$ to $[0,1]^3$; this is necessary for technical reasons,
since the GPU expects the coordinates in textures to be [0,1]. Our function $k(x,y,z)$ considers $(x,y,z)$ as azimuthal equidistant coordinates of a point
in $\bbH^3$ and returns its coordinates in the Poincar\'e ball model. Our choices of $i_x$, $i_z$ and $k$ ensure that the function $g(x,y,z)$ will be linear when
$x$ or $y$ are close to 1, and thus the interpolation will yield good approximate results for large values of $x,y,z$. 

\begin{figure}[ht]
\centering
\includegraphics[width=0.45\linewidth]{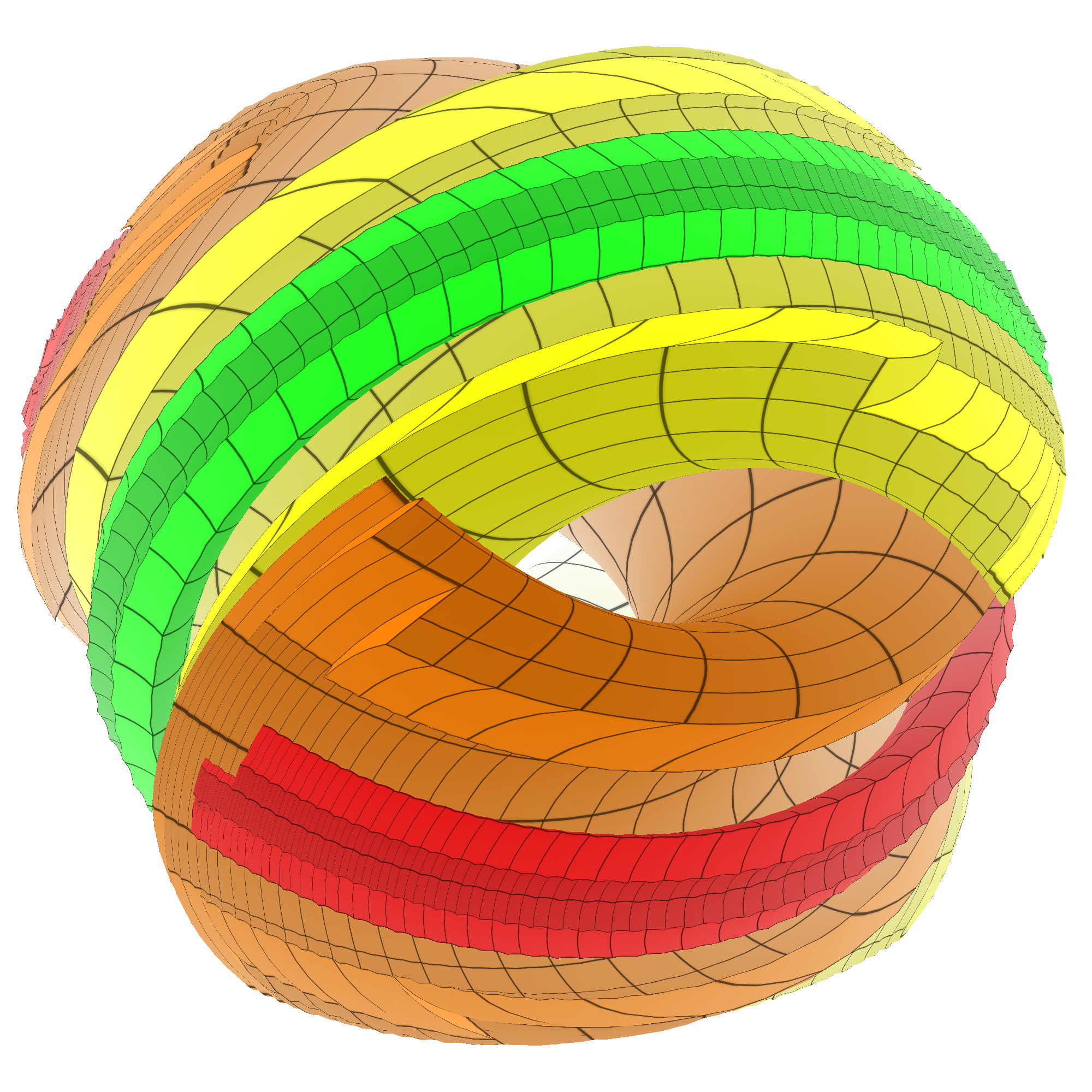}
\includegraphics[width=0.45\linewidth]{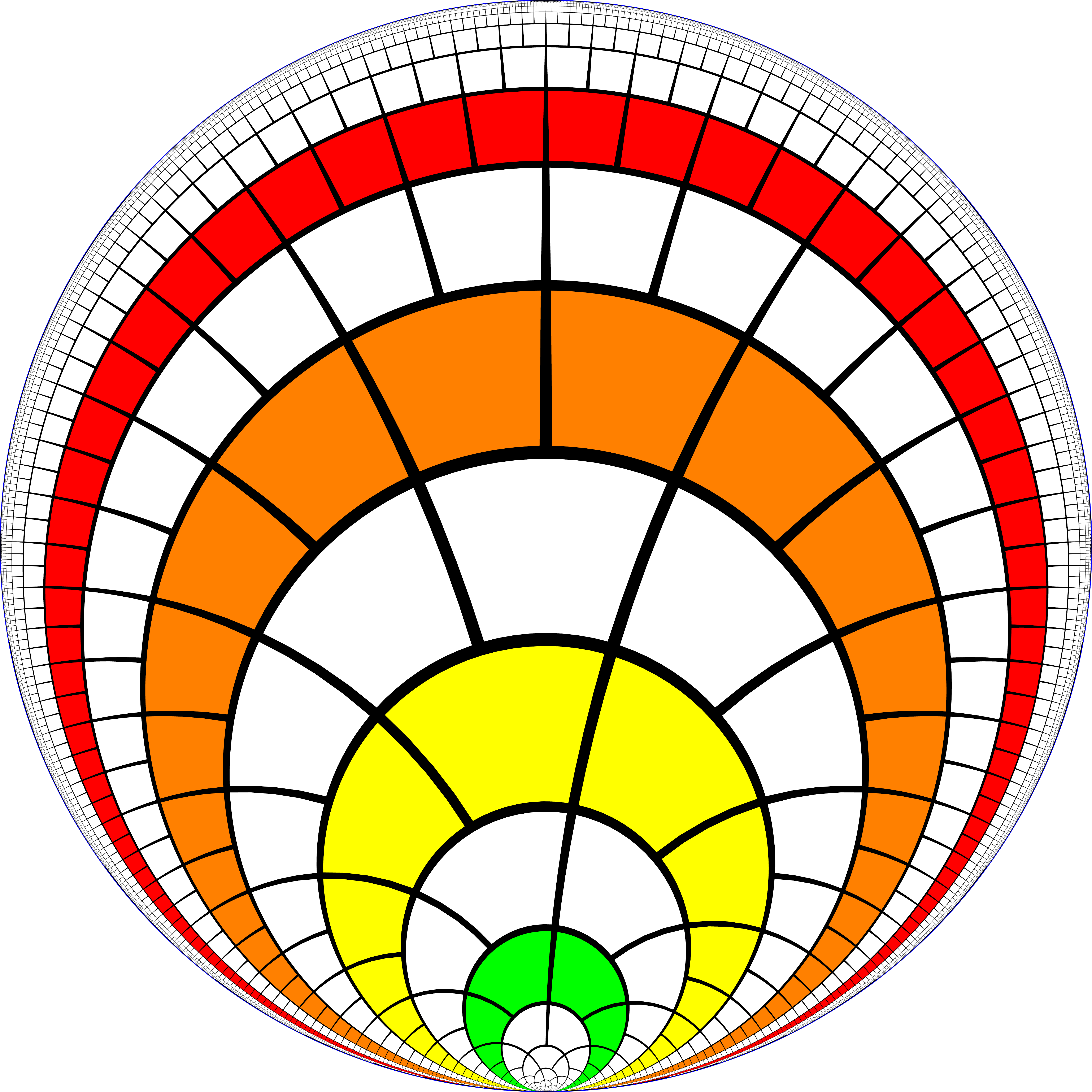}
\caption{$\Sol$ in a Poincar\'e-ball like model. From left to right: $\Sol$, 2D cut.\label{figball}}
\end{figure}

Figure \autoref{figball} illustrates $\Sol$ in a Poincar\'e-ball like model. The colored planes are surfaces of constant $z$. We graph $k(\log_0(x,y,z))$ for
the $\log_0$ computed using the method above. The cuts $x=0$ and $y=0$ are hyperbolic planes in the Poincar\'e disk model, as shown in \autoref{figball}b.
The surfaces of constant $z$ are mapped to torus-like shapes in this projection; the tori for $z>0$ and $z<0$ are interlocking. 
Since this model is azimuthal, it corresponds to what the user positioned in the center using a first-person perspective visualization perceives: surfaces
of constant $z$ are perceived as interlocking tori (\autoref{solvscene}).

The binary tiling we have used for the hyperbolic plane (Fig.~\autoref{figlet}) generalizes straightforwardly to {\bf Solv}. We will build the honeycomb 
in levels, where $i$-th level has the $z$ coordinate in range $((i-\frac{1}{2})\log 2,(i+\frac{1}{2})\log 2))$. On the level 0, our tessellation projects
to the tessellation of the plane $z=0$ by squares of side length $l$. Tessellation on level $i$ projects to rectangles of dimensions $1/{2^i} \times 2^i$. 
This way, we have subdivided Solv into isometric cube-like shapes. This tessellation can be implemented on a computer by representing every cell as a pair of
its projections to the hyperbolic planes $(x,z)$ and $(y,z)$; these projections are cells in the respective binary tilings on these planes.

Our method for computing $\log_0$ returns a single value. Therefore, the visualization based only on the direction computed with method outlined
so far does not
give the whole picture. However, it turns out that the effects of this issue are in fact minor. Multiple-valued $\log_0$ exist only for a small 
region of visible space (points $(x,y,z)$ where both $|x|$ and $|y|$ are greater than $\pi$, and $|z|$ is smaller than half of a single level of cells in our honeycomb),
and the geodesics that reach $a$ after multiple oscillations in $z$ tend to stretch objects in unrecognizable ways and to be hidden by other objects. Thus, our basic implementation 
does not handle this issue. The issue can be solved by using ray-based methods, using two-valued $\log_0$ which includes both
candidate geodesics computed (based on $p_1$ and $p_2$), or by special handling of the difficult case. Since $\log_0(x,y,z)$ computes the shortest geodesic,
it is not a continuous function in the points where the two geodesics computed from $p_1$ and $p_2$ are different paths of the same length (this happens
at points where $|x|=|y|>\pi$ and $z=0$); special care must be taken when rendering triangles that cross the non-continuous region.

\subsubsection{Geometries similar to {\bf Solv}}

Our approach can be used for other geometries similar to {\bf Solv}. Such geometries can be obtained by changing $g^{11}(x,y,z) = \exp(a_1z)$ and 
$g^{22}(x,y,z) = \exp(a_2z)$. For {\bf Solv} we have $a_1=1, a_2=-1$, for $\bbE^3$ we have $a_1=a_2=0$, for $\bbH^3$ we have $a_1=a_2=1$, for $\bbH^2\times\bbR$ we have
$a_1=1,\ a_2=0$. Thus, we can obtain a non-isotropic variant of hyperbolic space by taking for example $a_1 = \log 2$, $a_2 = \log 3$, or a less symmetric variant of {\bf Solv} 
by taking $a_1 = \log 2$, $a_2 = -\log 3$. Such geometries are not Thurston geometries, because there are no closed manifolds
which have these geometries. 

Our approach generalizes to such geometries. Our methods also generalize to $n$-dimensional versions of $\Sol$ for $n>3$, defined by 
$g^{ii}(x_1,\ldots,x_n) = \exp(a_ix_n)$ for $i=1, \ldots, n-1$ and $g^{nn}(x_1,\ldots,x_n) = 1$, $g^{ij} = 0$ for $i \neq j$. As long as the sequence $(a_i)$ contains only
two different non-zero values, we can compute $\log_0(x)$ using a three-dimensional precomputed texture by rotating $x$ so that it lies in a three-dimensional subspace.

\subsubsection{{\bf Nil} geometry}

Our implementation of {\bf Nil} geometry is based on the paper by Papp and Molnár \cite{nilgeodesics}. The translations of {\bf Nil} are given by the formula 
\begin{equation}M_(x,y,z)(a,b,c) = (x,y,z) * (a,b,c) = (a+x, b+y, c+xb+z) \label{nildef}\end{equation}

and thus can be represented as matrices when we add the fourth homogeneous coordinate.
The metric $g$ is given by $g(0)(v,w)$ is the inner product of $v$ and $w$, and for other points $a$, $g(a)$ is uniquely defined by the fact that $M_(x,y,z)$ is an isometry.

Formulas for Christoffel coefficients and geodesics in {\bf Nil} are computed in \cite{nilgeodesics}. It is also possible to determine the geodesics in Nil by gaining
enough geometric intuitions; these geometric intuitions will be also essential for us in the later sections.
In every point of Nil $p$ we have a local coordinate system, obtained by translating the local coordinate system at $(0,0,0)$ by $M_p$.
As a simple example, let us see what happens if we start in $(0,0,0)$, and move according to vectors $(d,0,0)$, $(0,d,0)$, $(-d,0,0)$, $(0,-d,0)$ (each according to the local coordinate
system in the point where we are). We start at $(0,0,0)$, then we get to $(0,0,0) * (d,0,0) = (d,0,0)$, then we get to $(d,0,0) * (0,d,0) = (d,d,d^2)$, $(d,d,d^2) * (-d,0,0) = (0,d,d^2)$,
and $(0,d,d^2)*(0,-d,0) = (0,0,d^2)$. Note that $d^2$ is the area of a square with edge $d$ which would be obtained if we took a similar path in Euclidean plane. 
An interesting consequence of this is that a path of length $d$ can change the $x$ and $y$ coordinates by up to $d$, but the $z$ coordinate can be changed up to $\Theta(d^2)$, and thus
the volume of a geodesic sphere in Nil is $\Theta(d^4)$.

In general, the geometrical intuition here is that every loop
$\gamma$ in the Euclidean space $\bbE^3$, starting at $(x,y,z)$, can be lifted to a path $\gamma'$ in Nil, which at every time $t$ traverses the space in the same local directions as $\gamma(t)$,
i.e., $\dot\gamma'(t) = M_{\gamma(t)} \dot\gamma(t)$. This path also starts at $(x,y,z)$, but ends at $(x,y,z+A)$, where $A = \int \gamma_x d\gamma_y(t)$.
Both paths are of the same length. By Green's theorem, this integral is the signed area inside the
projection of the loop $\gamma$ to the $XY$ plane, yielding a rotationally symmetric description of Nil geometry. This aspect makes {\bf Nil} similar to ``impossible figures'' such as Penrose triangles and Penrose staircases.

\begin{figure}[ht]
\begin{center}
\includegraphics[width=.45\linewidth]{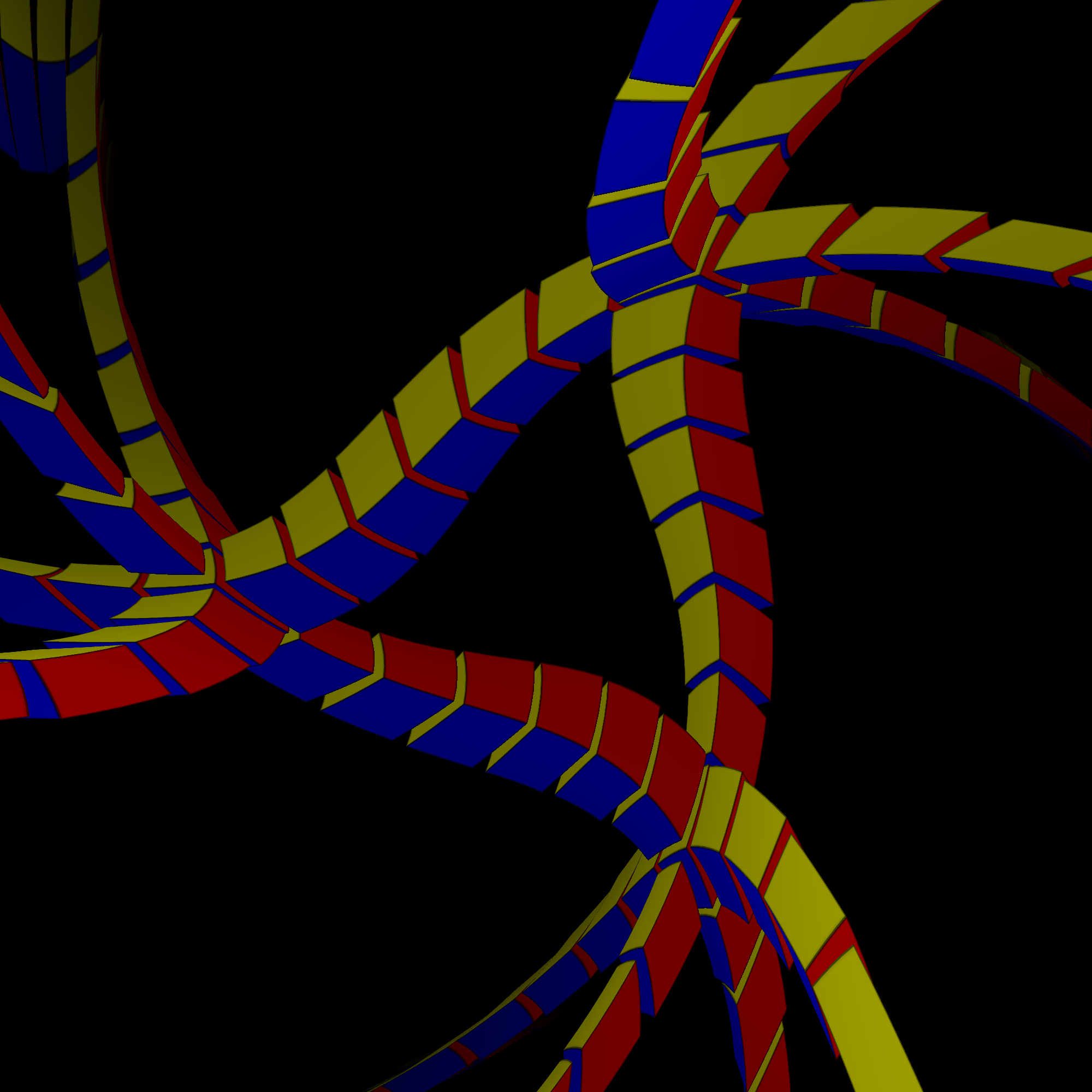}
\includegraphics[width=.45\linewidth]{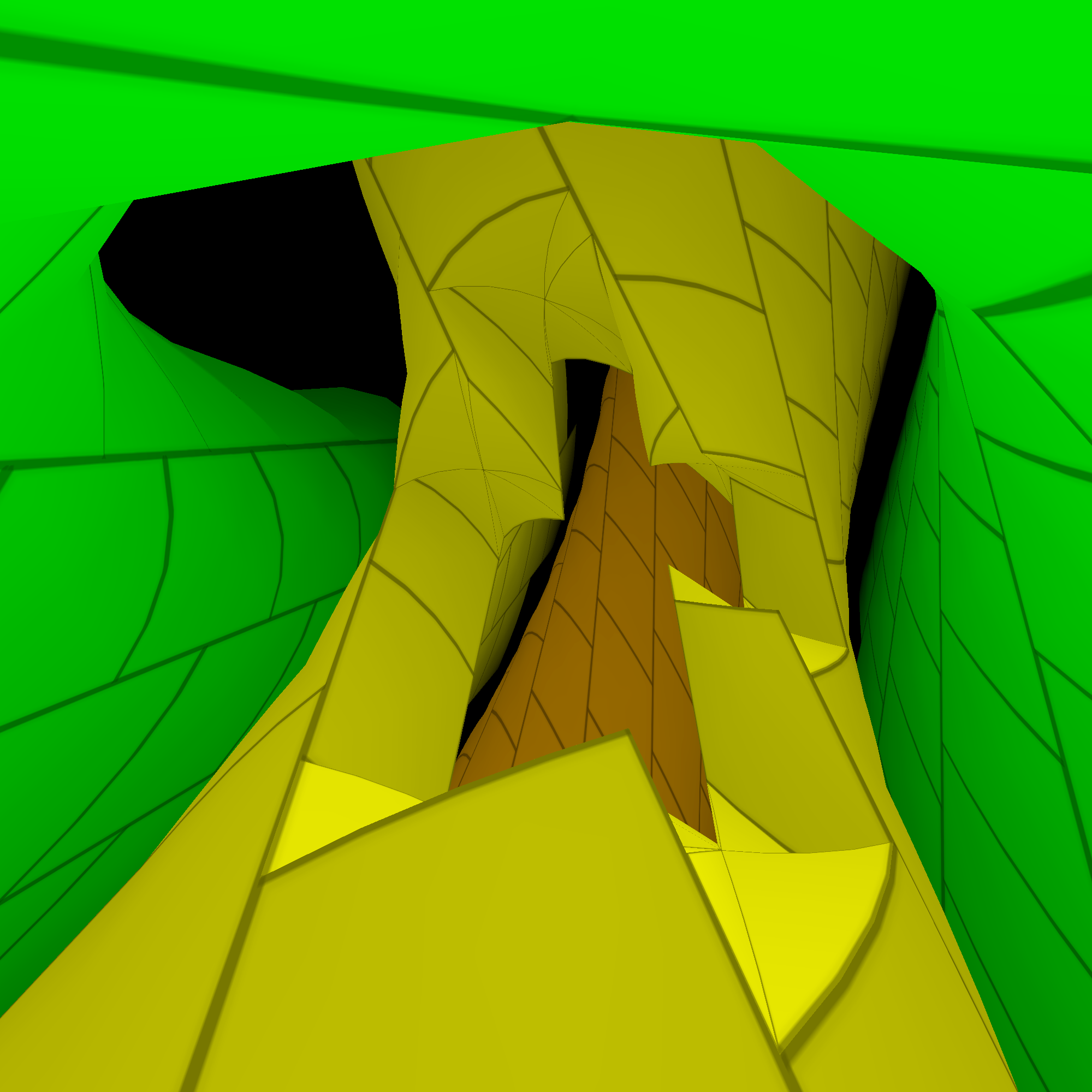}
\end{center}
\caption{A network of Penrose triangles in ${\bf Nil}$ geometry, and walls of cells of our honeycomb. \label{fignil}}
\end{figure}

It is easy to see that the example path that took us from $0$ to $d^2$ is not the shortest possible. Instead of lifting an Euclidean square with area $d^2$, we could also lift an 
Euclidean circle with area $d^2$. That circle has radius $r / \sqrt{\pi}$, and length $2r \sqrt{\pi} < 4r$. It is well known that a circle is the shortest curve which surronds a given
area; therefore, in general, the shortest paths will be projected to straight lines and circles in the $(x,y)$ plane. Consider a twisted cylindrical surface in Nil $(x,y,z): (x-r)^2+y^2=r^2$. 
This cylinder is glued in a twisted way: going around the cylinder changes the $z$ coordinate by $\pi r^2$. However, it has Euclidean geometry.
If we want to get from point $(0,0,0)$ to $(0,0,z)$ while looping once around the cylinder, the shortest path will be a straight line (helix) on this cylinder, which has length 
$\sqrt{(2\pi r)^2 + (z-\pi r^2)^2}$ by Pythagorean theorem. To find the shortest path from $(0,0,0)$ to $(0,0,z)$, we need to find $r$ which minimizes this value, which can be found by
differentiation. Since this is the shortest path, it will be a geodesic. Thus, the geodesics in Nil are either helices on the cylinders defined as above, or straight lines with tangent
vectors of form $M_v(x,y,0)$ at every $v$.

We use the formulas obtained in \cite{nilgeodesics}.
Let $t = (c \cos(\alpha), c \sin(\alpha), w)$. Then,
in the general case,
$\exp_0(t) = (\frac{c}{w} (\sin(wt+\alpha) - \sin(\alpha)), -\frac{c}{w} (\cos(wt+\alpha) - \cos(\alpha)), wt+\frac{c^2}{2w}t - \frac{c^2}{4w^2}(\sin(2wt+2\alpha)-\sin(2\alpha))
+ \frac{c^2}{2w^2}(\sin(wt+2\alpha)-\sin(2\alpha)-\sin(wt)))$. In the special case $c=0$ we have $\exp_0(t) = (0,0,wt)$ and in the case $w=0$ we have 
$\exp_0(t) = (ct \cos(\alpha), ct \sin(\alpha), \frac{1}{2} c^2 \cos(\alpha) \sin(\alpha) t^2)$. To find $t$ such that $\exp_0(t) = (x,y,z)$, note that, for the given value of 
$w$, we can compute $c$ and $\alpha$ that will give the correct $x$ and $y$, and then compute $z(w)$ based on $w$, $c(w)$, and $\alpha(w)$. The obtained $z(w)$, as a function of
$w \in (-2\pi, 2\pi)$, is monotonous, and thus we can find the correct $w$ using the bisection method. 
Other solutions exist where $|w| > 2\pi$, but these represent longer geodesics and similar to {\bf Solv} are less important in visualization.

The obvious honeycomb for {\bf Nil} has a cell for every point $(x,y,z) \in \bbZ^3$. The cells adjacent to $(x,y,z)$ are given by $(x,y,z) * \pm e_i$. The cells are not cubes -- after four moves $e_1$, $e_2$, $-e_1$, $-e_2$ we end up below the original cell. The side faces of cell 0 are given by $\pm\frac{1}{2}e_i * ke_j * le_3$, where
$k \in (-\frac{1}{2}, \frac{1}{2})$, $l \in (-\frac{1}{2}, \frac{1}{2})$, $i \in \{1,2\}$, $j=3-i$. The top and bottom face consists of four triangles given by $me_i * ke_j * (\pm \frac{1}{2}e_3)$,
where $m \in (-\frac{1}{2}, \frac{1}{2})$, $|k|<|m|$, $i \in \{1,2\}$, $j=3-i$, and four vertical walls connecting the four triangles. This construction makes the side faces of a cell
similar to that of a cube, while the top and bottom faces have little Penrose staircases on them. While these top and bottom faces are not flat, such a honeycomb is good for visualization,
as it shows the basic Penrose staircase-like nature of {\bf Nil}, as well as its rotational symmetry in the XY plane.

\subsubsection{Berger sphere}\label{sec:berg}

\begin{figure}[ht]
\begin{center}
\includegraphics[width=.45\linewidth]{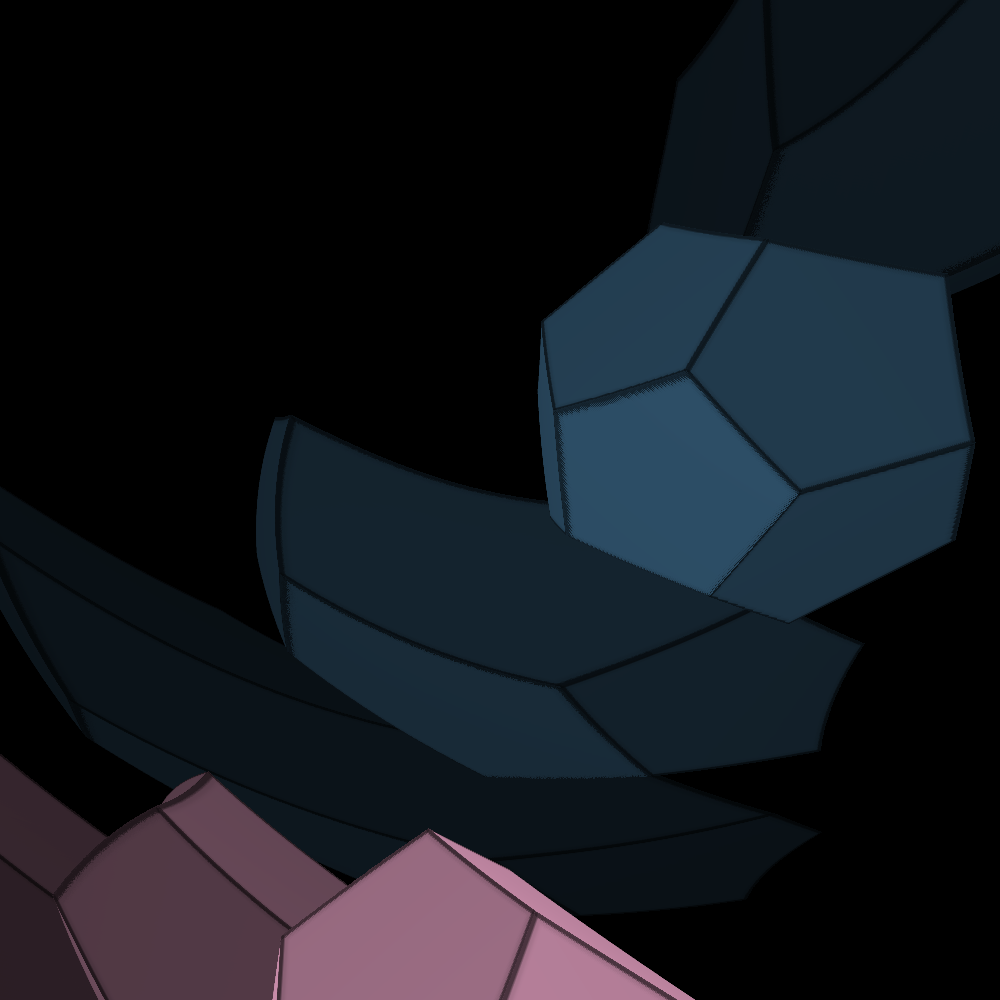}
\includegraphics[width=.45\linewidth]{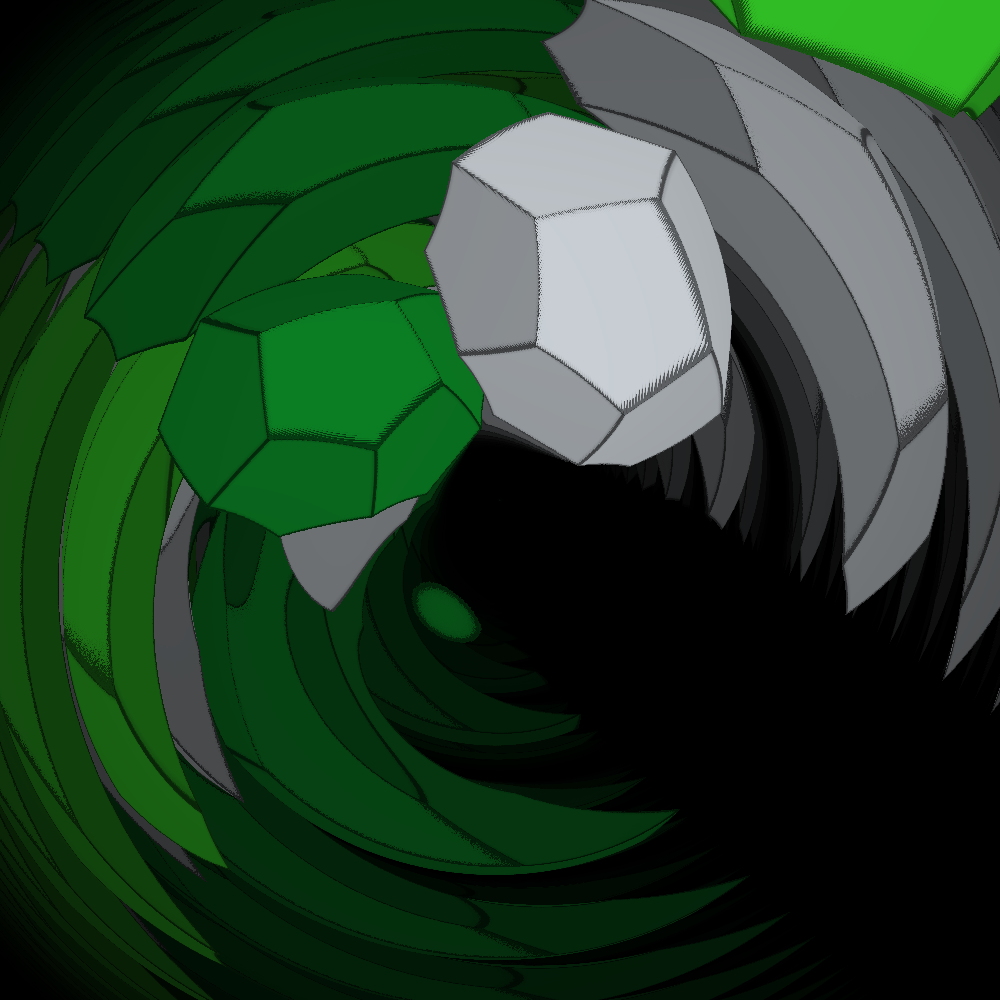}
\end{center}
\caption{Two scenes in the Berger sphere, $\alpha=0.9$ (left) and $0.5$ (right), obtained using raytracing.\label{sceneberg}}
\end{figure}

Let us repeat the geometric intuition we have obtained for {\bf Nil}: every loop
$\gamma$ in the Euclidean space $\bbE^3$, starting at $(x,y,z)$, can be lifted to a path $\gamma'$ in {\bf Nil}, which starts at $(x,y,z)$, but
ends at $(x,y,z+A)$, where $A$ is the signed area of the flat projection of the loop $\gamma$. {\bf Nil} geometry can be thus seen as ``twisted'' $\bbE^2\times \bbR$
\cite{weeks2001shape}; the same twisting operation can be also applied if the underlying two-dimensional space has $\bbS^2$ or $\bbH^2$ geometry. In this section, we will concentrate on 
twisted $\bbS^2\times\bbS^1$, i.e., the space where every loop $\gamma$ in $\bbS^2\times\bbR$ can be lifted to a path $\gamma'$ in $\bbS^2\times\bbR$, which
changes the value of the $z$ coordinate by the area of the projection of $\gamma$, multiplied by a factor $\alpha$. The factor $\alpha$ was not necessary in {\bf Nil} because its
only effect would be scaling the geometry; here, the choice of $\alpha$ is crucial. Since the areas of loops in $\bbS^2$ are only defined modulo $4\pi$, the third coordinate
is no longer the full line $\bbR$, but rather a circle $\bbS^1$ of length $4 \pi \alpha$.

However, with $\alpha=1$ our space has a special interpretation. 
As mentioned in Section \ref{tangentspace}, in $\bbS^2$ the parallel transport along a curve $\gamma'$ results in rotating vectors by angle equal to the area inside $\gamma'$.
Thus, our space becomes the space of rotations of $\bbS^2$, with the usual metric. This space of rotations (i.e., isometries of $\bbS^2$ that keep orientation) is the elliptic 3-space, i.e., the quotient space of $\bbS^3$
where we identify $(x,y,z,w)$ and $(-x,-y,-z,-w)$. This fact is frequently used in computer graphics, where unit quaternions are used for representing rotations of the three-dimensional
space. A point $v=(x,y,z,w) \in \bbS^3$ represents the following rotation of two-dimensional sphere given by the following matrix:

\begin{equation}
S_{(x,y,z,w)} = \left(
\begin{array}{ccc}
+x^2 - y^2 - z^2 + w^2 & -2(xy - zw) & 2(xz + yw) \\
-2(xy + zw) & -x^2 + y^2 - z^2 + w^2 & -2(yz - xw) \\
2(xz - yw) & -2(yz + xw) & -x^2 - y^2 + z^2 + w^2
\end{array}
\right).\label{qtm}
\end{equation}

We have $v * w = S_v * S_w$, where $v * w = M_v w$, where
\begin{equation}
M_{(x,y,z,w)} = \left(\begin{array}{cccc}w&-z&y&x\\z&w&-x&y\\-y&x&w&z\\-x&-y&-z&w\end{array}\right).\label{liesphere}
\end{equation}

Moving in the third coordinate in twisted $\bbS \times \bbR$ corresponds to rotating the sphere around the $z$ axis, which in turn corresponds to the tangent vector
$M_{(x,y,z,w)}(0,0,1,0)$. For $\alpha \neq 1$, we can use the same representation, but we need to stretch the metric along this tangent vector, i.e., 
for $v,w \in T_p(S^3)$, we define $<v,w> = <Z_\alpha M_{p}^{-1}(v), Z_\alpha M_{p}^{-1}(w)>$ where $Z_\alpha$ is a matrix which multiplies the third coordinate by $\alpha$.
This geometry is called Berger sphere \cite{berger_orig}. Berger sphere does not appear in Thurston's geometrization conjecture, because any compact manifold that
can be given this geometry can be also given the $\bbS^3$ metric, by stretching back along the fibers to make the geometry isotropic.

To find the geodesic formulas in the Berger sphere, we can use the same approach as we have been using for {\bf Nil}. While in {\bf Nil} we minimized
$\sqrt{(2\pi r)^2 + (z-\pi r^2)^2}$, in the Berger sphere we need to use the area and circumference formulas for spherical circles and account for $\alpha$; thus,
we minimize $\sqrt{(2\pi \sin(r))^2 + \alpha(z-2\pi (1-\cos(r))^2}$. (Here $z$ comes from the $\bbS^2\times\bbR$ representation, with $\alpha$-stretching is applied to it,
i.e., the $z$ coordinate repeats with period $4\pi$. The resulting spiral can be written as the following geodesic in the space of isometries of $\bbS^2$: $\Gamma_t = MR^{\alpha_0+t}X^rR^{-\alpha_0-t}R^{bt}$, where:
\begin{itemize}
\item $R^\alpha$ is rotation by $\alpha$ around the $z$ axis, $X^r$ is a spherical translation in the $X$ direction by $r$ units;
\item $R^{\alpha_0+t}X^rR^{-\alpha_0-t}$ is the movement along the circle, 
\item $R^{bt}$ is movement along the $z$ axis in $\bbS^2\times\bbS^1$, to turn the circular movement into a spiral; $b$ is the minimizing coefficient computed as above,
\item $M = R^{\alpha_0}X^{-r}R^{-\alpha_0}$ is to make $\Gamma_0$ equal identity.
\end{itemize}
Using (\ref{qtm}) and (\ref{liesphere}), we rewrite $\Gamma_t$ as a geodesic in the Berger sphere. We obtain the following formula: $\exp_M(x,y,z)$:

\begin{eqnarray*}
z'&=& z\alpha \\
l &=& \sqrt{x^2+y^2+z'^2} \\
z_0 &=& z'/l \\
x_0 &=& \sqrt{1-z_0^2} \\
\tan(r) &=& x_0/(z_0\alpha) \\
\tan(\beta) &=& y/x \\
z_1 &=& \cos(r) (1-1/\alpha^2) \\
a &=& l / \sqrt{\sin(r)^2 + (\cos(r)/\alpha)^2} \\
u &=& z_1a  \\
X &=& \sin(r) \sin(a) \cos(u+\beta) \\
Y &=& \sin(r) \sin(a) \sin(u+\beta) \\
Z &=& \cos(r) \sin(a) \cos(u) - \cos(a)\sin(u) \\
W &=& \cos(r) \sin(a) \sin(u) + \cos(a)\cos(u) \\
\exp_M(x,y,z) &=& (X,Y,Z,W)
\end{eqnarray*}

The angle $\beta$ should be in $[-\pi/2,\pi/2]$ if $z>0$ and in $-3\pi/2,-\pi/2$ otherwise.

We can compute all $(x,y,z)$ such that $\exp_M(x,y,z) = (X,Y,Z,W)$ as follows.
Let $A = \sqrt{X^2+Y^2}$. We have $\tan(r) = A/\sin(a)$. The argument
of $W+iZ$ is the following function of $a$: $\phi(a) = \arctan(\cos(r) \sin(a)/\cos(a)) - u$.
Therefore, we need to find the value of $a$ for which $\phi(a)$ agrees with the actual value of the argument (modulo $2\pi$).
Since $\tan(r) = A/\sin(a)$, we need $\sin(a) = A/\tan(r)$ be in the interval $[-1,1]$, therefore $a \in [2k\pi + \arcsin(A), 2k\pi+\pi-\arcsin(A)]$
for some $k$. For every $k$, the function $\phi(a)$ is monotonic or bitonic in the respective interval, thus we can find all the values 
by first finding the possible extremum (using ternary search) and then finding the actual values in both parts (using binary search), remembering
that $\phi(a)$ only has to agree with the known value modulo $2\pi$.
The value of $\beta$ only affects the rotation in the $XY$ coordinates, so after finding this value of $a$, the value of $\beta$ can be found easily.
Computing all possible $(x,y,z)$ is then straightforward.

Unfortunately, the compactness of Berger sphere, combined with it non-isotropic geometry, makes primitive-based rendering very difficult.
The algorithm above finds all the vectors where $(X,Y,Z,W)$ should be seen, but for a triangle with vertices $(v_1,v_2,v_3)$, it is not clear which point 
in $\exp_M^-1(v_1)$ should be matched with which point in $\exp_M^-1(v_2)$ and which point in $\exp_M^-1(v_3)$; also we should render many points if we
want to obtain a faithful representation.

Figure \ref{sceneberg} shows two scenes in Berger sphere. Because of the problems mentioned above, these pictured have been obtained using a raytracer
(see Section \ref{sec:ray} below). Our honeycomb here is based on the 120-cell in $\bbS^3$; the 120-cell can be arranged in
such a way that all the dodecahedra are aligned along the fibers (i.e., paths obtained by following the local $z$ coordinate), and every cell is obtained
from each other by a $S_v$ transformation for some $v$. Note that while the dodecahedra are regular in $\bbS^3$, in Berger sphere they are rotationally
symmetric, but no longer regular. In $\bbS^3$, the geodesics are great circles; in Berger sphere, the geodesics are instead helices around the fibers:
after making a $360^\circ$ loop, they do not get back to the point where they started, but a point slightly lower on higher on the fiber (for $\alpha$ close
to 1). For this reason, in Figure \ref{sceneberg}a we see multiple images of every dodecahedron (arranged along a fiber). With $\alpha$ further away
from 1, the scene becomes much more complex (Figure \ref{sceneberg}b): images of dodecahedra become ripped apart.

\subsubsection{Twisted $\bbH^2\times\bbR$}

\begin{figure}[ht]
\begin{center}
\includegraphics[width=.45\linewidth]{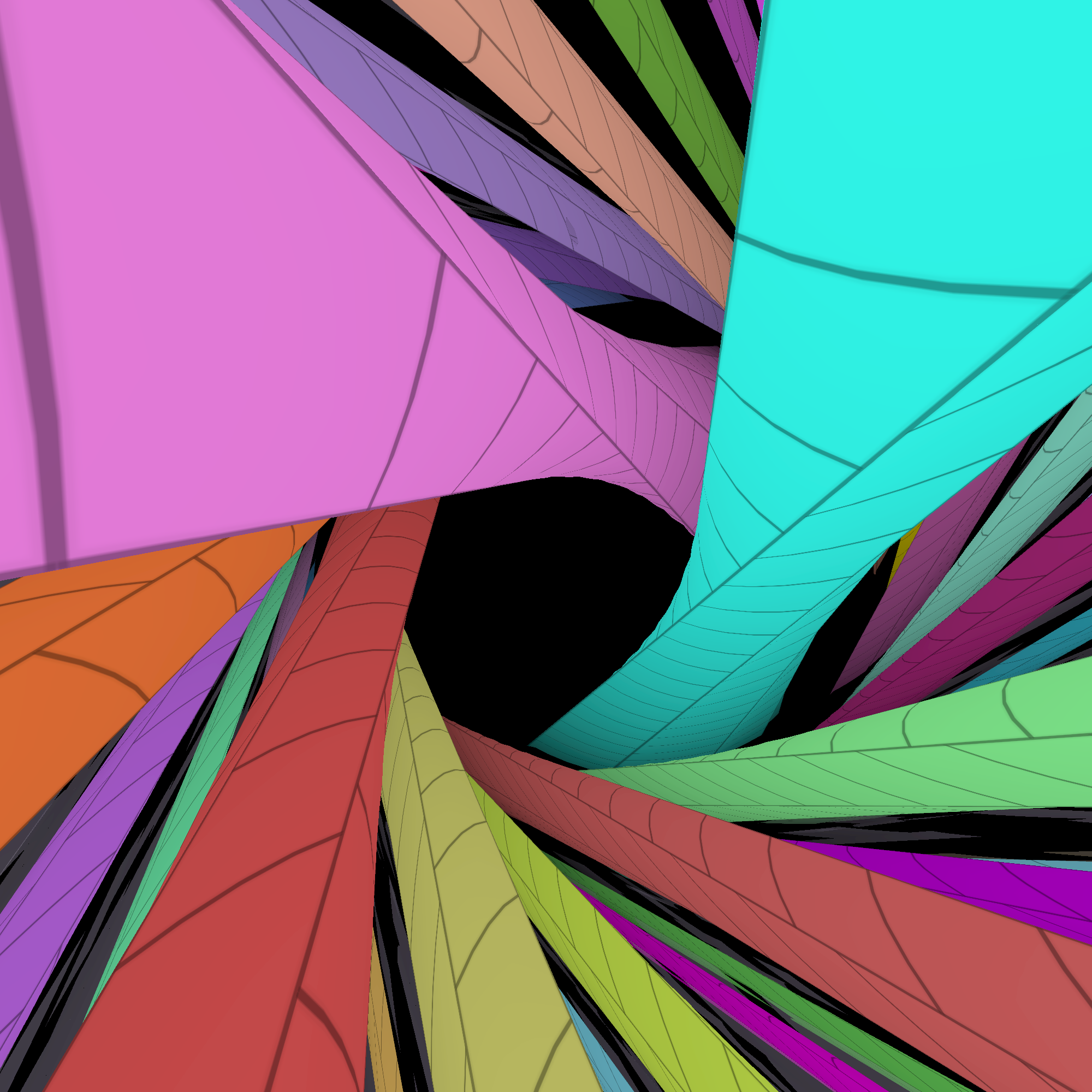}
\includegraphics[width=.45\linewidth]{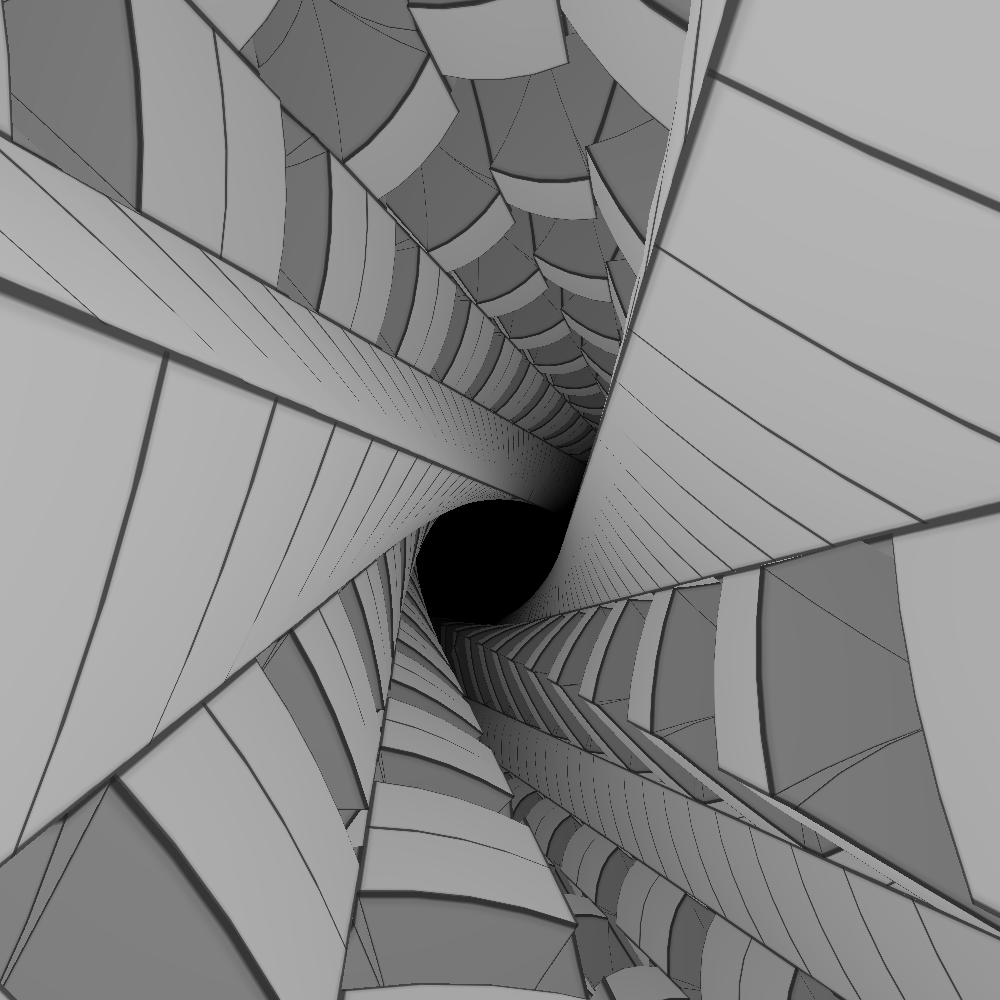}
\end{center}
\caption{Two scenes in the twisted $\bbH^2\times\bbR$ geometry. \label{scenepsl}}
\end{figure}

Similarly to the case of regular (non-twisted) $\bbH^2\times\bbR$ and $\bbS^2\times\bbR$, rendering twisted $\bbH^2\times\bbR$ turns out to be easier than rendering twisted $\bbS^2\times\bbS^1$.

We will render twisted $\bbH^2\times\bbR$ using similar approach to rendering $\bbS^2\times\bbS^1$. Again, the case $\alpha=1$ is the simplest and corresponds to the space of rotations of the hyperbolic plane.
We represent this space of rotations as the unit split-quaternions, i.e.,~$\{(x,y,z,w): z^2+w^2-x^2-y^2 = 1\}$. Just like in the case of $\bbS^2$, $v$ and $-v$ correspond to the same
rotation. We have

\[
S_{(x,y,z,w)} =
\left(
\begin{array}{ccc}
+x^2 - y^2 - z^2 + w^2 & -2(xy - zw) & -2(yz - xw) \\
-2(xy + zw) & -x^2 + y^2 - z^2 + w^2 & -2(xz + yw) \\
2(yz + xw) & 2(xz - yw) & x^2 + y^2 + z^2 + w^2
\end{array}\right),\]

and $v * w = S_v * S_w$, where $v * w = M_v w$, where
\begin{equation}
M_{(x,y,z,w)} = \left(\begin{array}{cccc}w&-z&y&x\\z&w&-x&y\\-y&x&w&z\\-x&-y&-z&w\end{array}\right).
\end{equation}

The geodesics can be computed in the same way as in $\bbS^2$, obtaining very similar formulas:

\begin{eqnarray*}
z'&=& z\alpha \\
l &=& \sqrt{x^2+y^2+z'^2} \\
z_0 &=& z'/l \\
x_0 &=& \sqrt{1-z_0^2} \\
\tanh(r) &=& x_0/(z_0\alpha) \\
\tan(\beta) &=& y/x \\
z_1 &=& \cosh(r) (-1-1/\alpha^2) \\
a &=& l / \sqrt{\sinh(r)^2 + (\cosh(r)/\alpha)^2} \\
u &=& z_1a  \\
X &=& \sinh(r) \sinh(a) \cos(u+\beta)\\
Y &=& \sinh(r) \sinh(a) \sin(u+\beta)\\
Z &=& \cosh(r) \sinh(-a) \cos(u) - \cosh(a)\sin(u) \\
W &=& \cosh(r) \sinh(-a) \sin(u) + \cosh(a)\cos(u) \\
\exp_M(x,y,z) &=& (X,Y,Z,W)
\end{eqnarray*}

If $|x_0/(z_0\alpha)| < 1$, this produces geodesics which project to circles in $\bbH^2$. 
In case if $|x_0/(z_0\alpha)| > 1$, these formulas also work, but $r$ has to be considered a
complex number. The formula produces geodesics which projects to equidistant curves or
straight lines in $\bbH^2$. 

The space we have obtained so far can be seen as twisted $\bbH^2\times\bbS^1$. Whether
we identify $v$ with $-v$ or not, this space is not simply connected; for example, the loop $(0,0,\sin(\phi),\cos(\phi))$, for $\phi \in [0,2\phi]$, is
not contractible. We can obtain a simply connected space (twisted $\bbH^2 \times \bbR$)
by adding $\phi \in \bbR$ as the fifth coordinate;
this $\phi$ is an argument of $iZ+W$. To take $\phi$ into account in our formula for $\exp$, we simply
set $\phi = \arctan(A \cosh(r)/(\sinh(r)\cosh(a))) - u$.

The geometry obtained is the last of the eight Thurston geometries; following \cite{thurston1982}, 
most sources call it the universal cover of $SL(2,\bbR)$. This name comes from the fact that
isometries of the hyperbolic plane can be identified with the group $PSL(2,\bbR)$, which is obtained
from the group $SL(2,\bbR)$ by identifying opposite points. These correspond to split quaternions
with antipodal points identified or not. Split quaternions and elements of $SL(2,\bbR)$ are 
very similar -- they both can be described as quartuples of real numbers, the difference is the
choice of base in $\bbR^4$ \cite{slgeodesics}. We strongly prefer the name twisted $\bbH^2\times\bbR$
\cite{weeks2001shape} for several reasons. First, $\bbH^2\times\bbR$ is already simple connected, no need to specify
the universal cover. Second, $SL(2,\bbR)$ suggests a specific representation; we prefer to use
a model-agnostic name, and for the internal model we prefer split quaternions,
because they are a direct analog of quaternions (just like the Minkowski hyperboloid is a direct
analog of the sphere, contrary to the half-plane model that the $SL(2,\bbR)$ representation is based on).

\begin{figure}[ht]
\begin{center}
\def\svgwidth{.2\textwidth}
\begingroup%
  \makeatletter%
  \providecommand\color[2][]{%
    \errmessage{(Inkscape) Color is used for the text in Inkscape, but the package 'color.sty' is not loaded}%
    \renewcommand\color[2][]{}%
  }%
  \providecommand\transparent[1]{%
    \errmessage{(Inkscape) Transparency is used (non-zero) for the text in Inkscape, but the package 'transparent.sty' is not loaded}%
    \renewcommand\transparent[1]{}%
  }%
  \providecommand\rotatebox[2]{#2}%
  \newcommand*\fsize{\dimexpr\f@size pt\relax}%
  \newcommand*\lineheight[1]{\fontsize{\fsize}{#1\fsize}\selectfont}%
  \ifx\svgwidth\undefined%
    \setlength{\unitlength}{384bp}%
    \ifx\svgscale\undefined%
      \relax%
    \else%
      \setlength{\unitlength}{\unitlength * \real{\svgscale}}%
    \fi%
  \else%
    \setlength{\unitlength}{\svgwidth}%
  \fi%
  \global\let\svgwidth\undefined%
  \global\let\svgscale\undefined%
  \makeatother%
  \begin{picture}(1,2)%
    \lineheight{1}%
    \setlength\tabcolsep{0pt}%
    \put(0,0){\includegraphics[width=\unitlength,page=1]{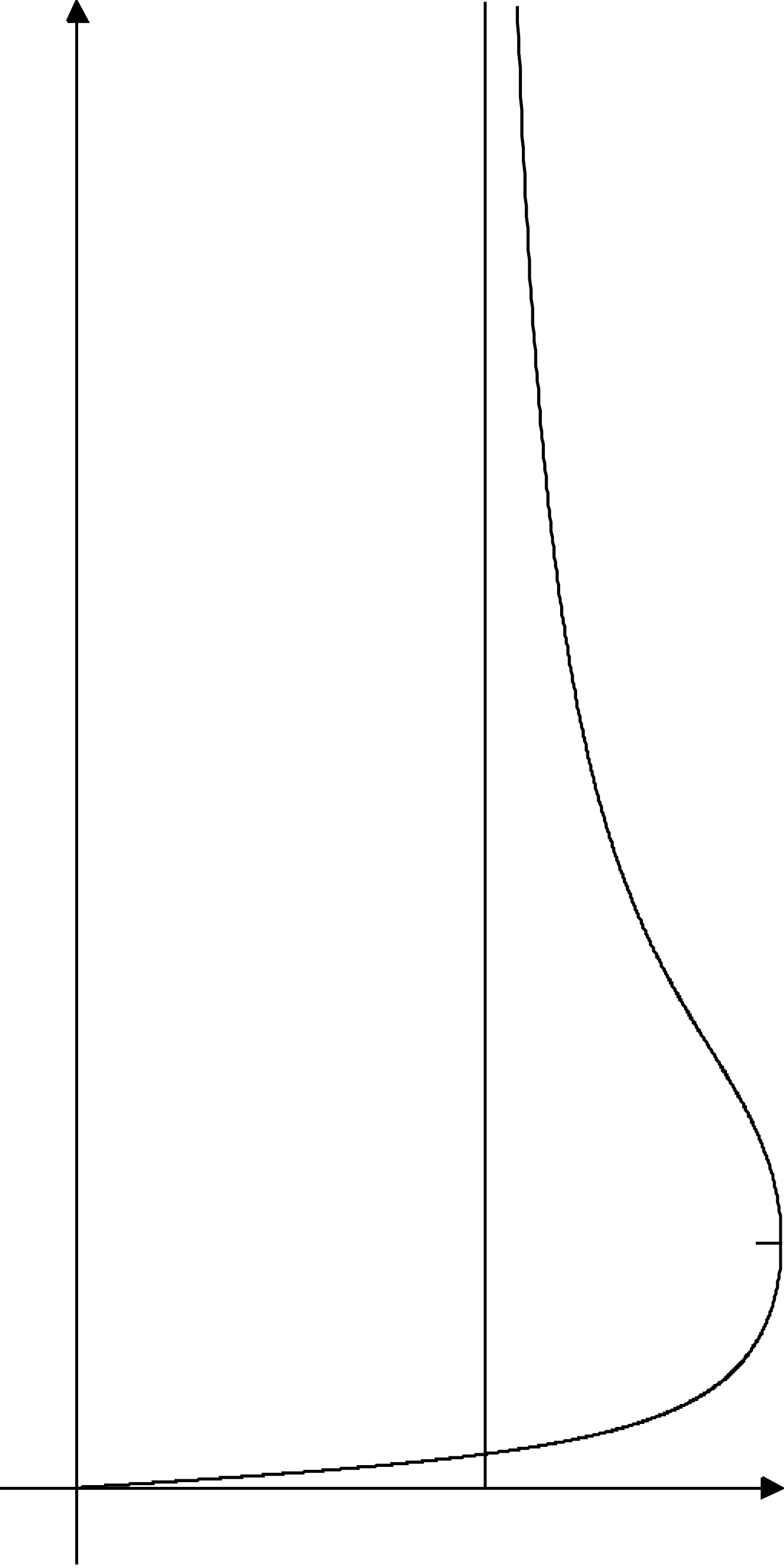}}%
    \put(0,2){\color[rgb]{0,0,0}\makebox(0,0)[lt]{\lineheight{1.25}\smash{\begin{tabular}[t]{l}$\phi$\end{tabular}}}}%
    \put(0.83788723,0.04030919){\color[rgb]{0,0,0}\makebox(0,0)[lt]{\lineheight{1.25}\smash{\begin{tabular}[t]{l}$s$\end{tabular}}}}%
    \put(0.55746341,0.16790605){\color[rgb]{0,0,0}\makebox(0,0)[lt]{\lineheight{1.25}\smash{\begin{tabular}[t]{l}$A$\end{tabular}}}}%
    \put(0.87839741,0.38793707){\color[rgb]{0,0,0}\makebox(0,0)[lt]{\lineheight{1.25}\smash{\begin{tabular}[t]{l}$B$\end{tabular}}}}%
  \end{picture}%
\endgroup%

\end{center}
\caption{The relationship between $\phi$ and $s$. \label{phisg}}
\end{figure}
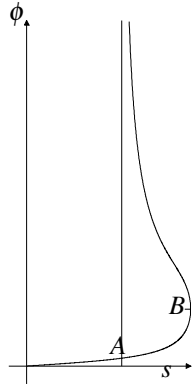

Even if the manifold we want to visualize is $\bbH^2\times\bbS^1$, we recommend thinking in terms of the universal cover
-- that is, to render a scene in $\bbH^2\times\bbS^1$, we render its counterpart in $\bbH^2\times\bbR$, in
which there will be a number of copies of every element in the original scene (number depending on the rendering
distance). Not only we get a more accurate rendering this way, but we also avoid the technical problems we had encountered while rendering looping spaces such as 
(twisted) $\bbS^2\times\bbS^1$, mentioned in Sections \ref{sec:prod} and \ref{sec:berg}.
To compute the inverse, assume that $\phi>0$ (the case $\phi<0$ can
be handled by flipping the $x$ and $z$ coordinates). We need to find $s \in [0,\pi/2]$ such that
$(x_0,z'_0) = (\cos(s), \sin(s))$. The relationship between $\phi$ and $s$ is shown in Figure \ref{phisg}.
For $s<\arctan(1/\alpha)$ (between the origin and the point A in Figure \ref{phisg}), we have the hyperbolic case where $r$ is not a real number; this case is 
simple. For greater $s$, we have $\sin(-a) = A / \sinh(r)$; if $A / \sinh(r) = 1$ we have
$-a = \pi/2$, otherwise we consider two possible values of $a$ (there are also other possible values,
but similarly to the {\bf Nil} case, they turn out to be not necessary for our visualizations). One of the possible
values is $\arcsin(A/\sinh(r))$ (from A to B in Figure \ref{phisg}), and the other is $\pi-\arcsin(A/\sinh(r))$
(from B upwards in Figure \ref{phisg}.

After computing the coordinates of the two points $A$ and $B$, we know in which of the section of
the graph on Figure \ref{phisg} should we look for the value of $s$ which yields the requested value of $\phi$.
The correct value of $s$ can then be found using binary search.
After finding $s$ and $a$, computing $(x,y,z)$ such that $\exp_M(x,y,z) = (X,Y,Z,W)$ is straightforward
($\beta$ can be computed just like for Berger sphere). 

Our earlier implementation of twisted $\bbH\times\bbR$ was based on Divjak et al. \cite{slgeodesics}. 
The formulas given by Divjak are of different form than ours. 
There is one small error in \cite{slgeodesics}: according to \cite{slgeodesics}, $\theta = \arctan(\sin(\alpha) \cdot \tan(s))$, 
but we should not take the principal value of $\arctan$, but rather the one such that the closest integer to $\theta/\pi$ should be the same as the closest integer to $s$.
After taking care of this issue, both formulas they yield the same result. Our formulas generalize to the
case $\alpha\neq 1$, and are also easier to work with, because they are not affected by this issue.

Our construction of a honeycomb in twisted $\bbH^2\times\bbR$ and twisted $\bbH^2\times\bbS^1$ is analogous to the construction of a honeycomb in {\bf Nil}, but based on a regular tessellation of the hyperbolic plane, 
instead of the square tessellation of the Euclidean plane. Just like in $\bbH^2 \times \bbR$, every cell is identified by
two generalized coordinates: the cell in the two-dimensional tessellation and the z-level. Contrary to $\bbH^2 \times \bbR$, the height of a single z-level
is not chosen arbitrarily, but corresponds to the smallest area of a polygon whose edges are line segments connecting the centers of cells of our tessellation.
If the $k$-th edge of a hyperbolic cell $c$ is the $k'$-th edge of the cell $c'$, then the $k$-th edge of a twisted cell $(c,z)$ is the $k'$-th edge of a twisted cell
$(c',z+\delta_{c,k})$; the values of $\delta_{c,k}$ have to be decided in such a way that, whenever we take a loop in $\bbH^2$, the respective path in the twisted space
needs to change the $z$-level proportionally to the area of the loop. This is straightforward for the space of rotations (i.e., $PSL(2,\bbR)$), as (assuming clockwise order
of cell neighbors) it is sufficient to just take $\delta_{c,k} = \pi + (2\pi/k) - (2\pi/k')$, where $2\pi$ is the full period. For $\bbH^2\times\bbR$, or $\bbH^2\times\bbS^1$
with any other period, we need to use the tree structure from Figure \ref{treegen}; for tree edges we have $\delta_{c,k}=0$, and for the other edges, $\delta_{c,k}$ corresponds to
the area of the loop in $\bbH^2$ constructed from the segment from $c$ to $c'$, and the unique path in the tree connecting $c$ and $c'$. Figure \ref{scenepsl} uses this
construction based on the \{7,3\} tessellation of the hyperbolic plane. Figure \ref{scenepsl}a is $PSL(2,\bbR)$ with $\alpha=1$, and the period in the $z$ coordinate is 14 levels.
Figure \ref{scenepsl}b uses the same tessellation, but with $\alpha=1/2$, and the period in the $z$ coordinate is 3 levels (thus, this is not a quotient space of $PSL(2,\bbR)$ nor
$SL(2,\bbR)$).

\section{Evaluation}

\subsection{Error analysis}

Our implementation of $\exp_M$ and $\log_M$ for {\bf Solv} and similar geometries is based on approximations. In this section, we study the errors introduced by our method for {\bf Solv}.
These errors arise for two reasons. 

\begin{figure}[ht]
\begin{center}
\includegraphics[width=\linewidth]{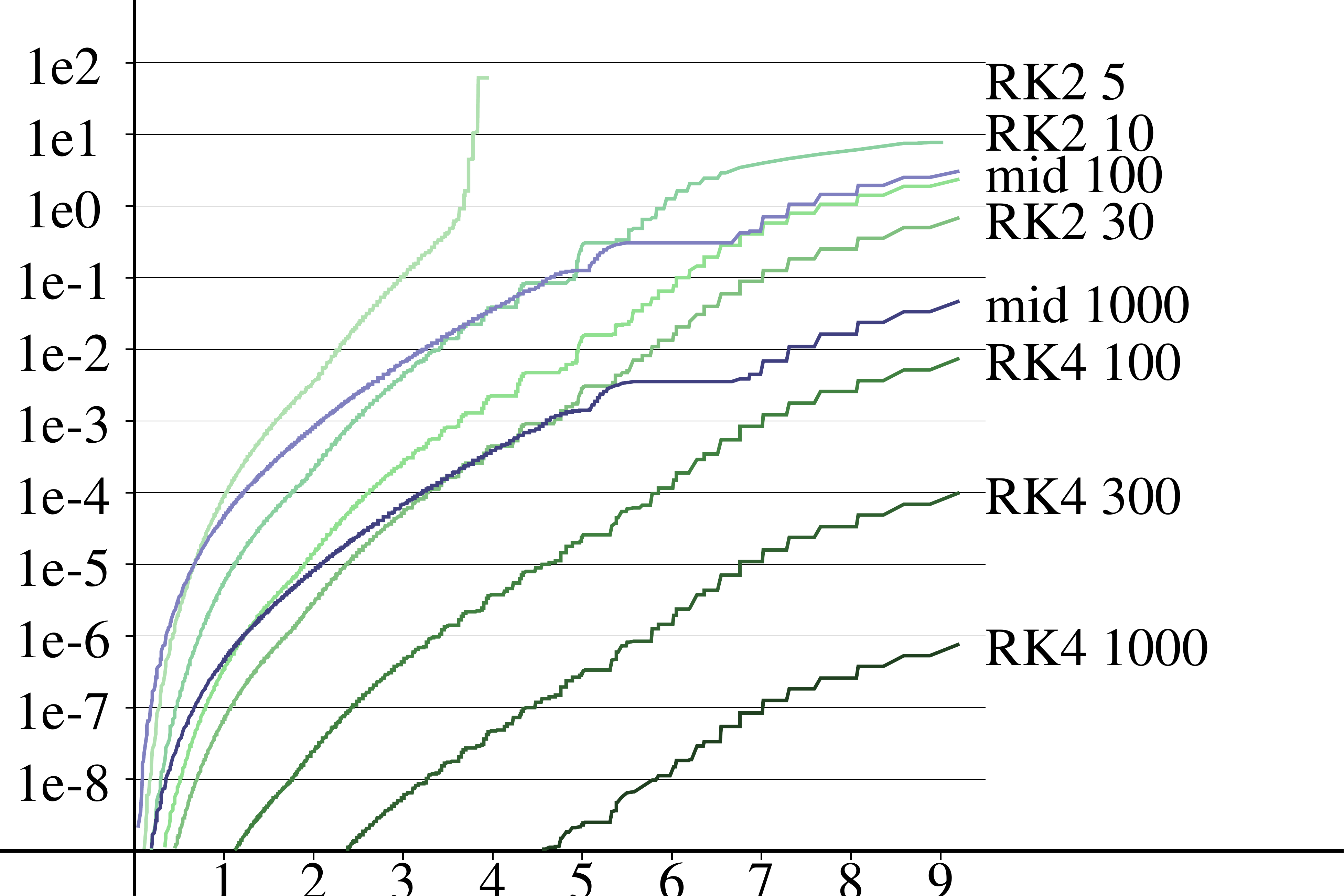}
\end{center}
\caption{Errors arising from solving differential equations numerically.}\label{errgraph}
\end{figure}

{\bf Solving differential equations numerically.} To compute $\exp_M$ we need to solve the geodesic equation. In Figure \autoref{errgraph}, we can see
the maximum distance between $\exp_M(v)$ and its approximation, as a function of $|v|$. As the correct value we use the result returned by RK4 method 
with 2000 steps. We can see that RK4 with 100 steps yields very good precision for the range we are interested in, and RK4 with 20 steps 
(unlabeled in Fig.~\autoref{errgraph}) quickly yields very good precision for a smaller range.

\begin{table*}[!ht]
  \begin{center}
 \caption{Errors arising from interpolation (in logarithmic scale).}\label{errtable}   
\small\setlength{\tabcolsep}{3pt}
\begin{tabular}{|c|cccc|cccc|cccc|cccc|}
\hline
 & \multicolumn{4}{|c}{$z_0$} & \multicolumn{4}{|c}{$z_1$} & \multicolumn{4}{|c}{$z_2$} & \multicolumn{4}{|c|}{$z_3$}\\
 & $x_0$ & $x_1$ & $x_2$ & $x_3$ & $x_0$ & $x_1$ & $x_2$ & $x_3$ & $x_0$ & $x_1$ & $x_2$ & $x_3$ & $x_0$ & $x_1$ & $x_2$ & $x_3$\\
\hline$y_0$  & -3.7 & -3.4 & -2.9 &   -2 & -3.2 & -3.1 & -2.8 & -2.1 & -2.8 & -2.8 & -2.5 &   -2 & -2.3 & -2.4 & -2.2 & -1.6\\
$y_1$  & -3.5 & -3.2 & -2.7 & -1.7 & -3.2 &   -3 & -2.7 & -1.9 & -2.8 & -2.7 & -2.5 & -1.9 & -2.3 & -2.4 & -2.2 & -1.5\\
$y_2$  & -2.9 & -2.7 &   -2 & -1.3 & -2.9 & -2.7 & -2.3 & -1.7 & -2.7 & -2.6 & -2.4 & -1.7 & -2.3 & -2.3 & -2.1 & -1.5\\
$y_3$  &   -2 & -1.5 & -1.2 & -0.6 &   -2 & -1.8 & -1.6 & -0.9 & -2.2 & -2.1 & -1.9 & -1.2 & -2.2 & -2.2 &   -2 & -1.3\\
\hline 
$y_0$  & -3.7 & -4.1 & -4.3 &   -4 & -3.9 & -4.3 & -4.6 &   -4 & -4.3 & -4.2 & -4.5 & -3.8 &   -4 & -3.9 & -4.3 & -3.5\\
$y_1$  & -3.9 & -4.1 & -4.1 & -3.6 & -3.8 &   -4 & -4.2 & -3.7 & -3.9 &   -4 & -4.3 & -3.8 & -3.9 & -3.9 & -4.2 & -3.5\\
$y_2$  & -4.1 & -3.8 & -3.6 & -3.1 &   -4 &   -4 & -3.9 & -3.4 & -3.9 & -3.9 & -4.1 & -3.6 & -3.8 & -3.8 & -4.1 & -3.5\\
$y_3$  &   -4 & -3.7 & -3.5 & -2.9 &   -4 & -3.8 & -3.7 & -3.1 & -3.8 & -3.7 & -3.5 & -3.3 & -3.7 & -3.7 & -3.8 & -3.3\\
\hline
\end{tabular}
\end{center}
\end{table*}

{\bf Error arising from interpolation.} Another source of errors is interpolation. We create a table of values of $\log_M(x,y,z)$
for $63^3$ points $(f_x(x), f_y(y), f_z(z))$, where $x, y, z \in \{0, \ldots, 62\}$ (63 corresponds to infinity), 
and we obtain $\log_M$ for other points by using linear interpolation. 
Table \autoref{errtable} presents the error introduced by this. We checked all the points $(f_x(x), f_y(y), f_z(z))$,
where $4x$, $4y$ and $4z$ are integers from $0$ to $239$. For every point checked, we computed the approximation $a = \log_M(x,y,z)$
using the method described and compared it with the actual $b = \log_M(x,y,z)$, obtained using the Newton method starting from $a$.
Since the approximation error depends on $x,y,z$, we split the values of coordinates into regions: the $i$-th region of $x$ ($i \in \{0,1,2\}$), denoted $x_i$ in the table, 
consists of $4x \in \{64i, \ldots, 64i+63\}$; region $x_3$ is a bit smaller, with $4x \in \{63\cdot 3, \ldots, 239\}$.
Regions are defined in the same way for $y$ and $z$. 
The interval boundaries correspond to actual values $f_x(x) \in \{0, 0.84, 2.05, 5.1, 26.7\}$ for $x$ and $y$,
and $f_z(z) \in \{0, 0.52, 1.12, 2, 3.71\}$.
For each of 64 regions, we give $log_{10}(|b|-|a|)$ in the top half of the table and the decimal logarithm of the angle between $a$ and $b$ in the bottom half. For each region,
the value given is the median error among all the checked points in this region.
We can see that angular errors are all on the order of $10^{-3}$ or better; this corresponds to one pixel when using $\log_M$ to render a first-person perspective projection.
A closer inspection of the errors reveals that such one pixel error is obtained for less than 2.5\% of points (the 97.5 percentile for error distribution is 0.001).
The errors in distances are larger, especially in the difficult region where both $x$ and $y$ are large while $z$ is small. Still, the distribution of the errors in distances is similar to the distribution of the errors in angles (a highly right-skewed distribution). However,
the errors here are irrelevant for first-person perspective visualizations that are our core interest.
 
\subsection{Validation using ray-based method}\label{sec:ray}
To validate the accuracy of the results of our primitive-based method, we have decided to compare them with
the results of ray marching.
We have implemented ray-based rendering for all the geometries.
To render each pixel, we send a ray starting in the current camera position in the direction depending on the pixel coordinates and camera
orientation. We find out where it hits a wall. Given a honeycomb in our manifold, we 
use coordinates relative to the cell $C$ the ray is currently in, and we need to find out which face $F$ of $C$ the ray hits. If the cell $C'$ on the other side of $F$ is filled,
the ray ends (and we color the pixel appropriately), otherwise we compute the coordinates relative to $C'$ and continue tracing the ray.

In the case of isotropic and product manifolds, it is straightforward to find formulas for the distance we need to travel in order to hit a plane $F$. In {\bf Solv} and its variants, 
we proceed by making small steps of length $u$ and compute the new coordinates after each such step (using the RK4 method). If it turns out that after $u$ units we are already
in another cell, we use bisection to find the collision point up to precision $\epsilon$. We halve $u$ and repeat as long as $u>\epsilon$. We have shown experimentally that
$u=0.05$ and $\epsilon = 0.001$ yield enough precision to be not readily distinguishable from more accurate computations. The similar method can be used in {\bf Nil} and other
twisted geometries, except that we can use precise geodesic formulas instead of the RK4 method, and therefore a larger maximum step value of $u=0.1$ works.

Our experiments show that primitive-based and ray-based methods yield the same output (modulo multiple geodesics), which shows that the approximations we have used when computed
$\log_0$ for {\bf Solv} indeed do not destroy the visualization effect. While multiple geodesics are not handled perfectly by primitive-based rendering, ray-based methods also
confirm that the differences are not significant (except in the Berger's sphere). As mentioned in Section \ref{homoco}, another verification is performed by checking whether the object seen
in the exact center of the screen remains in the center when the camera is moving forward. All three systems use implementations which are separate to some extent: camera movement
uses Christoffel symbols on CPU, raytracing uses either exponential functions or Christoffel symbols on GPU, and primitive-based rendering uses inverse exponential functions.

\section{Discussion and Applications}

In this section, we want to discuss the differences between our and competitive approaches. 
We also describe
areas that in our opinion benefit from our contribution.

{\bf Related works.}
Older visualizations of non-isotropic geometries include works by Weeks \cite{weeksreal} and Berger \cite{pierreberger}. Weeks
\cite{weeksreal} visualized
$\bbS^2 \times \bbR$ and $\bbH^2 \times \bbR$; plans for creating visualizations of other non-isotropic Thurston geometries were mentioned.
Our implementation of a primitive-based renderer for $\bbS^2 \times \bbR$ improves this work by
subdividing triangles close to the point directly above or below the camera or its antipodal point.
Berger \cite{pierreberger} included all Thurston geometries except twisted $\bbH^2\times\bbR$. However, those visuzalizations are static images rather than real-time rendered, which makes them difficult to interpret.
In late 2019 the subject has 
received attention of
three other teams \cite{magmamcfry,visgraf,trettelnil,trettelsol}. 
Novello et al. \cite{visgraf} are fully independent from our work, MagmaMcFry\cite{magmamcfry} is based on our early work.
All these approaches are based on ray-based algorithms. One aspect that is done in \cite{trettelnil,trettelsol} but mostly ignored by us is lighting. We have decided to not deal with
lighting in our implementation, because in exponentially expanding geometries such as $\bbH^3$, a single light will illuminate only a small volume of the space around it, and therefore
to create a realistic scene we would need to simulate a huge number of independent light sources; in particular, we cannot assume that the light comes from a infinitely far away sun, like
in Euclidean space. The problem can be solved in specific geometries and scenes, but there seems to be no good way to solve it in the scenes we find the most interesting.

{\bf Comparison with ray marching.} Competitive attempts are based on ray-based algorithms: for every pixel, they trace the ray (geodesic), and color the pixel depending on
the object that the ray hit. Our method is primitive-based: we represent our objects as triangles,
and we compute the screen position for every vertex. We have already showed that our method is at least not worse than ray marching (our experiments suggest there are no major visible differences in the placement of walls). Here we will discuss the advantages.

For basic visualization our method is more challenging. It is easier to trace geodesics than to find a geodesic
which hits the given point in a non-isotropic space. 
Solving the challenge of computing $\log_v$ efficiently is crucial for other reasons in our applications. 
For example, in a FPS video game, we need to compute which direction an enemy has to shoot to hit the player; non-Euclidean machine learning and 
physics simulations are usually computationally expensive and based on geodesic distances.
Our methods outperform ray marching in rendering shapes that are
generated in a more complex way, such as 3D models.
This makes them more applicable for gaming and scientific visualizations. 
Moreover, Virtual Reality relies on displaying separate
images for both eyes. 
When we see an object at a specific point in our single eye vision, this means that the object is on a line; our brain
then finds out where the lines defined by left eye and the right eye cross. This process, together with raytracing, works in isotropic geometries,
with only the minor disadvantage of incorrect depth perception.
The world is perceived as stretched Klein/gnomonic projection, which makes
$\bbH^3$ look bounded while $\bbS^3$ looks unbounded.
Non-isotropic ray-based VR will not work correctly, as the rays perceived by both eyes do not cross.
We could do non-isotropic primitive-based VR by finding out the direction and distance to every object and then using
an Euclidean renderer to render it in the right spot for both eyes. This makes our approach currently the only one
suitable for efficient VR applications with three-dimensional display. VR applications are going to be a subject of our future work.


{\bf Differences in motivation.} Implementations by low-dimensional topologists
mostly aimed at visualizing the compact manifolds 
and depicting local effects in the geometries, such as holonomy or lensing effects. Practical applications were not the focal point.
Our motivation is different: we want to work with large-scale structures
that are not necessarily periodic. Our methods combat problems resulting from {\it exponential growth} of negatively curved spaces,
where large-scale computations are
susceptible to floating-point errors \cite{reptradeoff}.
Our visualizations are based on tessellations, which are constructed precisely without using floating point arithmetics
and thus circumvent these problems. Tessellations are also used to build landmarks that can be used to navigate our
spaces, and are important by themselves in the applications in data analysis (e.g., the self-organizing maps \cite{ritter99,ontrup})
or in gaming (level design).

{\bf Applications.} In machine learning, a common approach is to embed data into a manifold, in such a way that the relationships between the points correspond to the relationships within our data. While Euclidean geometry is used most commonly, 
non-Euclidean geometries have recently proven useful: hyperbolic geometry \cite{papa} for hierarchical data and spherical \cite{sphereembedding,deepsphereembedding} and product \cite{productembedding} geometries 
for other data. We suppose our methods should facilitate working with non-isotropic geometries in data analysis; this will be a direction of our future work. 

Other than the scientific purposes, the visualization of non-isotropic geometries has potential
applications in video games or art. Many popular (mostly independent) video games experiment with spaces that work
differently from our Euclidean world. This includes spaces with weird topology
(Portal, Antichamber, Manifold Garden), interactions between 2D and 3D (Perspective, Fez, Monument Valley), non-Euclidean geometry
(HyperRogue), extra dimensions (Miegakure). Similar experimentation also happens in art. Such games and art are interesting 
not only for mathematicians and physicists wanting to understand these spaces intuitively, but also for casual players
curious to challenge their perception of the world. Non-isotropic geometries are especially relevant here because of their easily observable
weirdness. {\bf Nil}, a reminiscent of Penrose's staircases and M.~C.~Escher's artworks, should be promising for game design.


\section{Our implementation}

The methods described in this paper have been implemented as a part of our
non-Euclidean visualization engine, RogueViz \cite{hyperrogue,hrviz}. 
RogueViz 11.3x includes the following real-time-rendered visualizations: 

{\bf Snowballs.} 
We spread
balls randomly throughout the space, in such a way that the number of balls
in every subset $V$ is Poisson distributed, with expected value proportional to the
volume of $V$. Such a visualization is useful to show the properties of the space
itself (parallax effects, non-isotropy, expansion) rather than the properties of the
honeycomb used. The implementation still uses honeycombs as a way to organize the
snowballs. For $\Sol$ and $\Nil$ the snowball visualization provides a good way of checking
the visual quality of primitive-based rendering (e.g., by checking if there are any areas where it appears
that snowballs are missing, due to not rendering all solutions to $\log_M$).

{\bf Honeycombs.} In this part, we show the honeycombs we use in $\bbS^2\times\bbR$,
$\Sol$, $\Nil$, and twisted $\bbH^2\times\bbR$ geometries. Traditional visualizations of non-Euclidean geometries
\cite{weeksrealh,pierreberger,hyperbolicvr} show the edges of the honeycomb cells. This
method is often misleading, as our Euclidean brains tend to interpret such visualization
incorrectly. This is especially visible in the case of the \{4,3,5\} honeycomb,
where our brains assume that the angles of squares have 90 degrees, and only 4 fit
around every edge. As a result, in the case of non-isotropic
geometries it may be difficult to understand the structure shown. Therefore, we display
our honeycombs by filling some of the cells; cells are filled with different colors to
exhibit the structure better. This visualization works both with primitive-based method
and raycasting. Because of the exponential growth, the number of cells rendered is
very high, which makes the rendering distance is a bit low (for both methods).
This part also includes a 3D model of Fig.~\autoref{figball} and a scene specially designed
to exhibit the area in Solv that is difficult to render with primitive methods.

{\bf Impossible figures in Nil.} This part shows 3D models of impossible figures
(Penrose triangle and an impossible ring) rendered in $\Sol$.

HyperRogue also includes a simple racing game in nonisotropic geometries
(press Ctrl+T while in the start menu). 
Racing mode can be also turned off (in 'o') for a more random environment.
In the settings, 3D configuration can be used to enable or disable raycasting,
or to change the rendering distance. (Note that, especially outside of the racing mode,
the rendering distance may appear low -- however, the number of cells rendered
in this range is quite high because of exponential expansion.)

The source code (compilable under Linux) and Windows binaries are included. 
The following files are the most relevant for this paper:
\begin{itemize}
\item nonisotropic.cpp -- implementation of nonisotropic geometries
\item raycaster.cpp -- implementation of ray-based rendering
\item hyperpoint.cpp -- basic geometry routines for all geometries
\item devmods/solv-table.cpp -- producing the geodesic tables for Solv and its variants
\end{itemize}

Some visualization videos made using our engine:
\begin{itemize}
\item \url{https://youtu.be/C8HoCf_hkn8} -- a simple structure in {\Sol} geometry.
\item \url{https://youtu.be/2LotRqzibdM} -- a longer video in {\Sol}. This video uses an older version of our renderer; some of its details are different.
\item \url{https://youtu.be/YmFDd49WsrY} -- a Penrose triangle in {\Nil} geometry.
\item \url{https://youtu.be/3WejR74o6II} -- impossible ring in {\Nil} geometry.
\item \url{https://youtu.be/HeFyuVs-Tts} -- our honeycomb in {\Nil} geometry.
\item \url{https://youtu.be/2ePY7Do5WvA} -- a structure in {\PSL}.
\item \url{https://youtu.be/_5l8v6Gn2sE} -- a structure in {$\bbS^2\times\bbR$} geometry.
\item \url{https://youtu.be/Hg-IW6XfgZY} -- a structure in twisted $\bbH^2\times\bbR$ ($\alpha=0.5$).
\item \url{https://youtu.be/zwRKzaE_7Mo} -- Berger sphere (ray-based rendering).
\item \url{https://youtu.be/KBYPQaoBgz0} -- another video in Berger sphere (ray-based rendering).
\item \url{https://youtu.be/leuleS9SpiA} -- Snowball visualization in video form.
\end{itemize}

\section{Conclusions}
Non-isotropic geometries are of great interest in low-dimensional topology and have potential applications
in cosmology, data analysis or game design.
In this paper, we presented novel methods of real-time native geodesics rendering of first-person
perspective in non-isotropic three-dimensional geometries. 
The greatest technical challenge we overcome
is computing the inverse exponential function in the $\Sol$ and similar geometries.
Our approach, based on primitives and tessellations, is currently the only one suitable for large-scale visualizations, visualization of complex scenes,
or VR applications with three-dimensional display. 
 Our computational methods can be 
also applied to machine learning and video games.

{\bf Acknowledgments.} 
We would like to thank the HyperRogue community, in particular to Kaida Tong and MagmaMcFry for sparking our interest in the {\Sol} geometry
and discussions. We would also like to thank the organizers of the ICERM Illustrating Geometry and Topology Workshop, 
partially funded by the Alfred P. Sloan Foundation award G-2019-11406 and supported by a Simons Foundation Targeted Grant to Institutes,
for inviting us; this workshop has been a big inspiration for our work. We would also like to thank Jeff Weeks for useful suggestions which
have improved this article, and Craig Hodgson for discussions. This work has been supported by the National Science Centre, Poland, grant UMO-2019//35/B/ST6/04456.


\def\ext#1{#1}
\bibliography{../master}

\begin{thebibliography}{10}

\bibitem{berger_orig}
M.~Berger.
\newblock Les vari\'et\'es riemanniennes homog\`enes normales simplement
  connexes \`a courbure strictement positive.
\newblock {\em Annali della Scuola Normale Superiore di Pisa - Classe di
  Scienze}, 3e s{\'e}rie, 15(3):179--246, 1961.

\bibitem{pierreberger}
P.~Berger.
\newblock Espaces imaginaires, 2015.
\newblock
  \url{http://www.espaces-imaginaires.fr/works/ExpoEspacesImaginaires2.html}
  (accessed Jan 12, 2020).

\bibitem{solvgeodesics}
A.~Bölcskei and B.~Szilágyi.
\newblock Frenet formulas and geodesics in sol geometry.
\newblock {\em Beiträge zur Algebra und Geometrie}, 48(2):411--421, jan 2007.

\bibitem{boroczky}
K.~Böröczky.
\newblock Gömbkitöltések állandó görbületű terekben {I}.
\newblock {\em Matematikai Lapok}, 25:265--306, 1974.

\bibitem{cannon}
J.~W. Cannon, W.~J. Floyd, R.~Kenyon, Walter, and R.~Parry.
\newblock Hyperbolic geometry.
\newblock In {\em In Flavors of geometry}, pp. 59--115. University Press, 1997.
\newblock Available online at
  \url{http://www.msri.org/communications/books/Book31/files/cannon.pdf}.

\bibitem{hrviz}
D.~Celi\'nska and E.~Kopczy\'nski.
\newblock Programming languages in github: {A} visualization in hyperbolic
  plane.
\newblock In {\em \ext{Proceedings of the Eleventh International Conference on
  Web and Social Media,} {ICWSM}, Montr{\'{e}}al, \ext{Qu{\'{e}}bec,} Canada,
  May 15-18, 2017.}, pp. 727--728. The AAAI Press, Palo Alto, California, 2017.

\bibitem{trettelnil}
R.~Coulon, E.~A. Matsumoto, H.~Segerman, and S.~Trettel.
\newblock Non-euclidean virtual reality {III}: Nil, 2020.
\newblock arXiv 2002.00513.

\bibitem{trettelsol}
R.~Coulon, E.~A. Matsumoto, H.~Segerman, and S.~Trettel.
\newblock Non-euclidean virtual reality {IV}: Sol, 2020.
\newblock arXiv 2002.00369.

\bibitem{slgeodesics}
B.~Divjak, Z.~Erjavec, B.~Szabolcs, and B.~Szil{\'a}gyi.
\newblock Geodesics and geodesic spheres in {SL(2 ; R)} geometry.
\newblock {\em Mathematical Communications}, 14(2):413--424, dec 2009.

\bibitem{thurston_physics}
J.~Gegenberg, S.~Vaidya, and J.~F. V{\'a}zquez-Poritz.
\newblock Thurston geometries from eleven dimensions.
\newblock {\em Classical and Quantum Gravity}, 19(23):L199--L204, nov 2002.
  doi: {{%
10\hspace{.1pt}\discretionary{.}{%
}{.}\hspace{.4pt}1088\discretionary{/}{%
}{/}0264\discretionary{%
}{-}{-}9381\discretionary{/}{%
}{/}19\discretionary{/}{%
}{/}23\discretionary{/}{%
}{/}102}}


\bibitem{productembedding}
A.~Gu, F.~Sala, B.~Gunel, and C.~R{\'e}.
\newblock Learning mixed-curvature representations in product spaces.
\newblock In {\em Proc.\ ICLR}, pp. 1--21. OpenReview.net, 2019.

\bibitem{hyperbolicvr}
V.~Hart, A.~Hawksley, E.~A. Matsumoto, and H.~Segerman.
\newblock Non-euclidean virtual reality {I}: explorations of $\mathbb{H}^3$.
\newblock In {\em Proceedings of Bridges: Mathematics, Music, Art,
  Architecture, Culture}, pp. 33--40. Tessellations Publishing, Phoenix,
  Arizona, 2017.

\bibitem{hyperrogue}
E.~Kopczy\'{n}ski, D.~Celi\'{n}ska, and M.~\v{C}trn\'{a}ct.
\newblock Hyper{R}ogue: Playing with hyperbolic geometry.
\newblock In {\em Proceedings of Bridges \ext{: Mathematics, Art, Music,
  Architecture, Education, Culture}}, pp. 9--16. Tessellations Publishing,
  Phoenix, Arizona, 2017.

\bibitem{trigridold}
E.~Kopczyński and D.~Celińska-Kopczyńska.
\newblock Hyperbolic triangulations and discrete random graphs, 2017.
\newblock (The current paper extends the theoretical part of the ArXiv paper;
  the experimental part is a subject of a further paper.).

\bibitem{deepsphereembedding}
W.~{Liu}, Y.~{Wen}, Z.~{Yu}, M.~{Li}, B.~{Raj}, and L.~{Song}.
\newblock Sphereface: Deep hypersphere embedding for face recognition.
\newblock In {\em Proc.\ CVPR}, pp. 6738--6746. IEEE, New York City, USA, 2017.
  doi: {{%
10\hspace{.1pt}\discretionary{.}{%
}{.}\hspace{.4pt}1109\discretionary{/}{%
}{/}CVPR\hspace{.1pt}\discretionary{.}{%
}{.}\hspace{.4pt}2017\hspace{.1pt}\discretionary{.}{%
}{.}\hspace{.4pt}713}}


\bibitem{ggtbook}
C.~L{\"o}h.
\newblock {\em Geometric Group Theory: An Introduction}.
\newblock Universitext. Springer International Publishing, 2017.

\bibitem{magmamcfry}
MagmaMcFry.
\newblock Solvview, 2019.
\newblock \url{https://github.com/MagmaMcFry/SolvView} (accessed Feb 6,2020).

\bibitem{margenstern_heptagrid}
M.~Margenstern.
\newblock Pentagrid and heptagrid: the fibonacci technique and group theory.
\newblock {\em Journal of Automata, Languages and Combinatorics},
  19(1-4):201--212, 2014. doi: {{%
10\hspace{.1pt}\discretionary{.}{%
}{.}\hspace{.4pt}25596\discretionary{/}{%
}{/}jalc\discretionary{%
}{-}{-}2014\discretionary{%
}{-}{-}201}}


\bibitem{visgraf}
T.~Novello, V.~da~Silva, and L.~Velho.
\newblock Visualization of {N}il, {SL2} and {S}ol, 2019.
\newblock \url{https://www.visgraf.impa.br/ray-vr/?page\_id=252} (accessed Feb
  6,2020).

\bibitem{ontrup}
J.~Ontrup and H.~Ritter.
\newblock Hyperbolic {S}elf-{O}rganizing {M}aps for {S}emantic {N}avigation.
\newblock In {\em Proc.\ NIPS}, pp. 1417--1424. MIT Press, Cambridge, MA, USA,
  2001.

\bibitem{papa}
F.~Papadopoulos, M.~Kitsak, M.~A. Serrano, M.~Bogu\~n\'a, and D.~Krioukov.
\newblock {Popularity versus Similarity in Growing Networks}.
\newblock {\em Nature}, 489:537--540, Sep 2012.

\bibitem{nilgeodesics}
D.~Papp and E.~Molnár.
\newblock Visualization of {N}il-geometry; modelling {N}il-geometry in
  euclidean space with software presentation, 2003.
\newblock \url{https://doi.org/10.25596/jalc-2014-201} (accessed Jan 12, 2020).

\bibitem{perelman}
G.~Perelman.
\newblock {The Entropy formula for the Ricci flow and its geometric
  applications}, 2006.
\newblock arXiv math/0211159.

\bibitem{gunnvis}
M.~Phillips and C.~Gunn.
\newblock Visualizing hyperbolic space: Unusual uses of 4x4 matrices.
\newblock In {\em Proc.\ I3D}, pp. 209--214. Association for Computing
  Machinery, New York, NY, USA, 1992. doi: {{%
10\hspace{.1pt}\discretionary{.}{%
}{.}\hspace{.4pt}1145\discretionary{/}{%
}{/}147156\hspace{.1pt}\discretionary{.}{%
}{.}\hspace{.4pt}147206}}


\bibitem{ritter99}
H.~Ritter.
\newblock Self-organizing maps on non-euclidean spaces.
\newblock In E.~Oja and S.~Kaski, eds., {\em Kohonen Maps}, pp. 97--108.
  Elsevier, 1999.

\bibitem{reptradeoff}
F.~Sala, C.~De~Sa, A.~Gu, and C.~Re.
\newblock Representation tradeoffs for hyperbolic embeddings.
\newblock In {\em Proc.\ ICML}, pp. 4460--4469. PMLR, Stockholmsm{\"a}ssan,
  Stockholm Sweden, 2018.

\bibitem{thurston1982}
W.~P. Thurston.
\newblock Three dimensional manifolds, {K}leinian groups and hyperbolic
  geometry.
\newblock {\em Bulletin (New Series) of the American Mathematical Society},
  6(3):357--381, may 1982.

\bibitem{weeks2001shape}
J.~Weeks.
\newblock {\em The Shape of Space}.
\newblock Chapman \& Hall/CRC Pure and Applied Mathematics. Taylor \& Francis,
  2001.

\bibitem{weeksrealh}
J.~Weeks.
\newblock Real-time rendering in curved spaces.
\newblock {\em IEEE Computer Graphics and Applications}, 22(6):90--99, Nov.
  2002. doi: {{%
10\hspace{.1pt}\discretionary{.}{%
}{.}\hspace{.4pt}1109\discretionary{/}{%
}{/}MCG\hspace{.1pt}\discretionary{.}{%
}{.}\hspace{.4pt}2002\hspace{.1pt}\discretionary{.}{%
}{.}\hspace{.4pt}1046633}}


\bibitem{weeksreal}
J.~Weeks.
\newblock Real-time animation in hyperbolic, spherical, and product geometries.
\newblock In M.~E. Prékopa~A., ed., {\em Non-Euclidean Geometries. Mathematics
  and Its Applications}, vol. 581, pp. 287--305. Springer, Boston, MA, jan
  2006. doi: {{%
10\hspace{.1pt}\discretionary{.}{%
}{.}\hspace{.4pt}1007\discretionary{/}{%
}{/}0\discretionary{%
}{-}{-}387\discretionary{%
}{-}{-}29555\discretionary{%
}{-}{-}0\_15}}


\bibitem{sphereembedding}
R.~C. {Wilson}, E.~R. {Hancock}, E.~{Pekalska}, and R.~P.~W. {Duin}.
\newblock Spherical and hyperbolic embeddings of data.
\newblock {\em IEEE Transactions on Pattern Analysis and Machine Intelligence},
  36(11):2255--2269, nov 2014. doi: {{%
10\hspace{.1pt}\discretionary{.}{%
}{.}\hspace{.4pt}1109\discretionary{/}{%
}{/}TPAMI\hspace{.1pt}\discretionary{.}{%
}{.}\hspace{.4pt}2014\hspace{.1pt}\discretionary{.}{%
}{.}\hspace{.4pt}2316836}}


\end{thebibliography}
\end{document}